\documentclass[iop,numberedappendix,twocolappendix]{emulateapj}

\usepackage[backref,breaklinks,colorlinks,citecolor=blue]{hyperref}
\usepackage[all]{hypcap}

\slugcomment{Submitted to Astrophysical Journal}
\shorttitle{Resolving the Discrepancy of Merger Fraction Measurements at $z \sim 0 - 3$}
\shortauthors{Man, Zirm \& Toft}

\begin{document}

\title{Resolving the Discrepancy of Galaxy Merger Fraction Measurements at $z \sim 0 - 3$}

\author{Allison W. S. Man, Andrew W. Zirm, and Sune Toft}
\affil{Dark Cosmology Centre, Niels Bohr Institute, University of Copenhagen, Denmark}
\email{allison@dark-cosmology.dk}


\begin{abstract}
We measure the merger fraction of massive galaxies using the UltraVISTA/COSMOS \textit{Ks}-band selected catalog, 
complemented with the deeper, higher resolution 3DHST+CANDELS catalog selected in the \textit{HST}/WFC3 \textit{H}-band,
presenting the largest mass-complete photometric merger sample up to $z\sim3$.
We find that selecting mergers using the $H_{160}$-band flux ratio leads to an increasing merger fraction with redshift, 
while selecting mergers using the stellar mass ratio causes a diminishing redshift dependence. 
Defining major and minor mergers as having stellar mass ratios of 1:1 - 4:1 and 4:1 - 10:1 respectively,
the results imply $\sim$1 major and $\lesssim$1 minor merger for an average massive (log$(M_{\star}/M_{\odot}) \geqslant 10.8$) galaxy during $z=0.1-2.5$.
There may be an additional $\sim 0.5(0.3)$ major (minor) merger if we use the $H$-band flux ratio selection.
The observed amount of major merging alone is sufficient to explain the observed number density evolution for the very massive (log$(M_{\star}/M_{\odot}) \geqslant 11.1$) galaxies.
We argue that these very massive galaxies can put on a maximum of $6\%$ of stellar mass in addition to major and minor merging, 
so that their number density evolution remains consistent with observations.
The observed number of major and minor mergers can increase the size of a massive quiescent galaxy by a factor of two at most.
This amount of merging is enough to bring the compact quiescent galaxies formed at $z>2$ to lie at $1\sigma$ below the mean of the stellar mass-size relation as measured in some works \citep[e.g.][]{Newman2012},
but additional mechanisms are needed to fully explain the evolution,
and to be consistent with works suggesting stronger evolution \citep[e.g.][]{vanderWel2014}.
\end{abstract}



\section{Introduction}

For decades, galaxy merging has been a popular explanation for the observed evolution in galaxy properties.
Galaxy mergers were first invoked to explain the morphological transformation of galaxies \citep{Toomre1972, Barnes1996}.
Merging remains the backbone in cosmological simulations in building up large galaxies \citep[e.g.][]{Springel2005, Bower2006}.
Gas-rich major mergers at high redshifts ($z>2$) are thought to trigger starburst and active galactic nuclei (AGN) episodes, quench star formation, and lead to bulge formation,
thereby building the massive ellipticals in the local Universe \citep{Barnes1991, Mihos1994, Kartaltepe2010, Toft2014}.
An alternative scenario has been proposed more recently,
in which massive galaxies at high redshift are clumpy disks which are very efficient in turning incoming cold gas into stars \citep{Dekel2009}.
The most luminous AGNs and ultra-luminous infrared galaxies (ULIRGs) take place in major galaxy mergers \citep{Kartaltepe2010, Treister2012, Ellison2013}.
Merging galaxies have enhanced star formation activity compared to isolated ones (\citealt{Patton2011, Yuan2012, Patton2013}; but also see \citealt{Xu2012b, Lanz2013}).
As galaxy merging may have profound influence on how the galaxy population evolved to this day, 
quantifying its rate of occurrence is essential to judge whether it explains any of the observed evolutionary trends.

As the timescale for galaxy mergers is on the order of a Gyr \citep[e.g.][]{Lotz2010a},
the conventional way to measure galaxy merger rate is to divide the observed fraction of galaxies undergoing mergers by a typical merging (observability) timescale at different redshift bins.
Merging galaxies can be identified as close galaxy pairs or galaxies displaying disturbed morphologies,
and the timescale required to convert the merger fraction to merger rate depends on the specific selection technique.
In this work we use the pair selection method,
as the merger fraction measured from morphological selection \citep[e.g.][]{LSanjuan2009, Bluck2012} and the merging observability timescale are dependent on the imaging depth and resolution.
The advent of multi-wavelength blank field observations in the past decade have enabled many improvements in the measurement of merger fractions, including the following: 
(1) the merger fraction of massive galaxies can be measured beyond $z\sim1$;
(2) the photometric redshifts allow more accurate removal of the pairs projected along the line-of-sight;
(3) the stellar masses derived from the spectral energy distribution (SED) fitting provide the stellar mass ratio of galaxy pairs, 
which is a more physically meaningful proxy for the dynamical interaction than a single-band flux ratio;
(4) deeper and wider area surveys provide larger samples, 
which in turn allow the dependence of merger fractions on different parameters to be explored.
Multiple authors have measured the merger fraction at $z>1$,
presenting somewhat conflicting results:
does the merger fraction increase with redshift \citep{Bluck2009, Man2012}, 
remain constant, or even diminish \citep{Williams2011, Newman2012}?
As shown in \citet{Lotz2011}, the variation of the parent galaxy selection and and mass ratio limits can contribute to some of the discrepancies across studies.
On the other hand, the average merging observability timescale is hard to estimate due to the large possible variety of orbital parameters and viewing angles,
as well as the lack of observed dynamical information on a galaxy-to-galaxy basis.
The uncertainties in the implied merger rates are discussed thoroughly in \citet{Hopkins2010b}.

In this work, 
we present the largest sample of photometrically selected mergers at $z$=0.1-3 to date from stellar mass complete catalogs.
The \textit{Ks}-band selected catalog from the UltraVISTA/COSMOS survey \citep{Muzzin2013} covers a large area, allowing us to expand our merger sample to more than five times times larger than previous studies.
We complement the ground-based UltraVISTA catalog with the space-based 3DHST+CANDELS \citep{Skelton2014} catalog,
which is deeper and has higher spatial resolution, 
to study possible systematic effects in measuring merger fractions.
The remainder of the paper is structured as follows:
Section~\ref{sec:data} describes the UltraVISTA and the 3DHST+CANDELS catalogs used in our study.
We present the criteria for selecting massive galaxies and mergers,
as well as the completeness of the catalogs.
In Section~\ref{sec:method} we present the method of measuring the merger fractions as a function of redshift.
We compare the merger fractions measured using the two catalogs, 
as well as the selection using the stellar mass ratio and $H_{160}$-band flux ratio.
We examine the stellar mass ratio distribution of the selected mergers.
We discuss the two main sources of uncertainties in the merger fraction measurements.
We show that we are complete to detecting minor mergers up to $z=2.5$.
Finally we convert the merger fractions to merger rates,
and infer the merger contribution in the stellar mass, size, velocity dispersion and number density evolution of massive galaxies.
Based on our findings, we address some broader questions in the context of galaxy evolution in Section~\ref{sec:discussion}:
What do the merger rates imply for the evolution of massive quiescent galaxies?
Is merging an influential process in the cosmic star formation history or not?
We also discuss the future prospects of merger fraction studies.
The conclusions of this work is summarised in Section~\ref{sec:conclusions}.
In Appendix~\ref{sec:missing_mergers} we present the simulations we perform to test for the completeness limits of the faintest possible satellites.
Appendix~\ref{sec:pf_lit} provides an in-depth comparison to similar merger fraction measurements in the literature.

All magnitudes are quoted in the AB system.
A cosmology of $H_{0}$ = 70 km s$^{-1}$ Mpc$^{-1}$, 
$\Omega_\mathrm{M}$ = 0.3 and $\Omega_{\Lambda}$ = 0.7 is adopted throughout this work.


\section{Data and sample selection} \label{sec:data}

\subsection{UltraVISTA catalog} \label{sec:uvista_catalog}
We use the \textit{Ks}-band selected catalog for the UltraVISTA Survey compiled by \citet{Muzzin2013}.
The UltraVISTA survey targets the COSMOS field \citep{Scoville2007} with the ESO VISTA survey telescope.
The effective survey area of UltraVISTA is 1.62 deg$^{2}$.
The catalog contains PSF-matched photometry in 30 photometric bands covering the wavelength range 0.15 - 24$\micron$ and includes the \textit{GALEX} \citep{Martin2005}, CFHT/Subaru \citep{Capak2007}, UltraVISTA \citep{McCracken2012}, S-COSMOS \citep{Sanders2007}, and zCOSMOS \citep{Lilly2007} datasets.
The UltraVISTA source detection is performed on the \textit{Ks}-band image with a $2.1\arcsec$ aperture, 
which has a limiting magnitude of $23.7 \pm 0.1$ ($5\sigma, 2\arcsec$-aperture).
In total there are 154 803 detected sources with reliable photometry having \textit{Ks}$<23.4$, which is the 90\% completeness limit and the adopted luminosity limit in this work. 
The stellar masses quoted in this paper are derived assuming a Chabrier IMF.
Further details regarding the photometric redshifts (photo-$z$'s) and SED fitting can be found in \citet{Muzzin2013}.

\subsection{3DHST+CANDELS catalog} \label{sec:3dhst_catalog}
To complement the ground-based \textit{YJHK$_{s}$} imaging from VISTA,
we use the 3DHST catalog presented in \citet{Brammer2012} and \citet{Skelton2014},
which includes \textit{HST} imaging from the CANDELS survey \citep{Grogin2011, Koekemoer2011} over five fields: COSMOS, GOODS-North and South, AEGIS, and UDS with a combined usable area of $\sim 0.25$ deg$^{2}$.
\citet{Skelton2014} performed photometry (aperture of $0.7\arcsec$) on the PSF matched images and compiled a photometric catalog with photo-$z$'s and SED best fits. 
We only use the objects marked with good photometry to ensure reliable photo-$z$'s and stellar masses.

\subsection{Selecting massive galaxies and mergers} \label{sec:select_pairs}

We use close galaxy pairs as a probe for galaxy mergers following similar criteria used in the literature \citep{Bluck2009, Williams2011, Man2012, Newman2012}.
In the UltraVISTA catalog, 
there are 9829 massive (log$(M_{\star}/M_{\odot}) \geqslant 10.8$) galaxies in the redshift range of $0.1 < z \leqslant 3.0$,
and 380 ($\sim 3.9\%$) of them are covered by the \textit{HST}/WFC3 \textit{H}-band imaging from the CANDELS and 3DHST COSMOS surveys.
Around these massive galaxies, we search for galaxy satellites fulfiling the following criteria:
\begin{enumerate}
	\item Within a projected separation of $R_{proj} = 10-30$ kpc $h^{-1}$.
	\item Stellar mass ratio $\mu=M_{1}/M_{2}$ of 1:1 - 4:1 as major merger, 4:1 - 10:1 as minor merger.
	\item The 1$\sigma$ confidence intervals of the photo-$z$'s  of the pair overlap.\end{enumerate}
We calculate $R_{proj}$ using the angular scale based on the photo-$z$'s of the more massive galaxy. 
As the FWHM of the ground-based UltraVISTA $Ks$-band image is $\sim0.8\arcsec$, corresponding to a maximum of 9.7 kpc $h^{-1}$ at $z\sim1.5$,
we use 10 kpc $h^{-1}$ as the lower limit of $R_{proj}$ to ensure that no close pairs are missed due to blending.
In Section~\ref{sec:compare_merger_rates} we explore the use of different $R_{proj}$ bins up to 100 kpc $h^{-1}$.
We explore the use of the $H$-band flux ratio as a probe for the stellar mass ratio in Section~\ref{sec:pf_ratios},
which we demonstrate to have a profound impact on the merger fraction evolution at $z>1.5$.
The redshift distribution of massive galaxies and pairs are listed on Table~\ref{table:pf_massratio}.

\subsection{Completeness limits}

We assess the completeness limit of the massive galaxies and their 4:1 and 10:1 satellites in two aspects:
the stellar mass completeness and the surface brightness limits.
We detail our analysis in Appendix~\ref{sec:missing_mergers} and give the summary as follows.
We find that the surface brightness limit is the constraining factor for detecting the satellites of massive galaxies.
If completeness is only estimated by comparing the magnitude-redshift distribution to deeper catalogs,
the completeness limits may be overstated.
We find that UltraVISTA (3DHST+CANDELS) is complete to $z=2.4$ and $z=1.5$ ($z=3.0$ and $z=2.5$) for major and minor mergers respectively.
In this work,
the data points at redshift bins which are mass incomplete are either omitted or plotted as lower limits,
to ensure that incompleteness does not affect our conclusions.
Despite the fact that 3DHST+CANDELS is deeper than UltraVISTA and can probe the merger fractions to higher redshifts,
we demonstrate in Section~\ref{sec:pf_uvista} that we do not get a higher merger fraction, both major and minor, with 3DHST+CANDELS compared to UltraVISTA,
suggesting that there is not a significant population of mergers that have faint quiescent satellites only detectable in the 3DHST+CANDELS catalog.


\begin{figure*}[htb]
	\begin{minipage}[b]{0.52\linewidth}
	\label{fig:pf_massratio}
	\centering
	\includegraphics[angle=0,width=\textwidth]{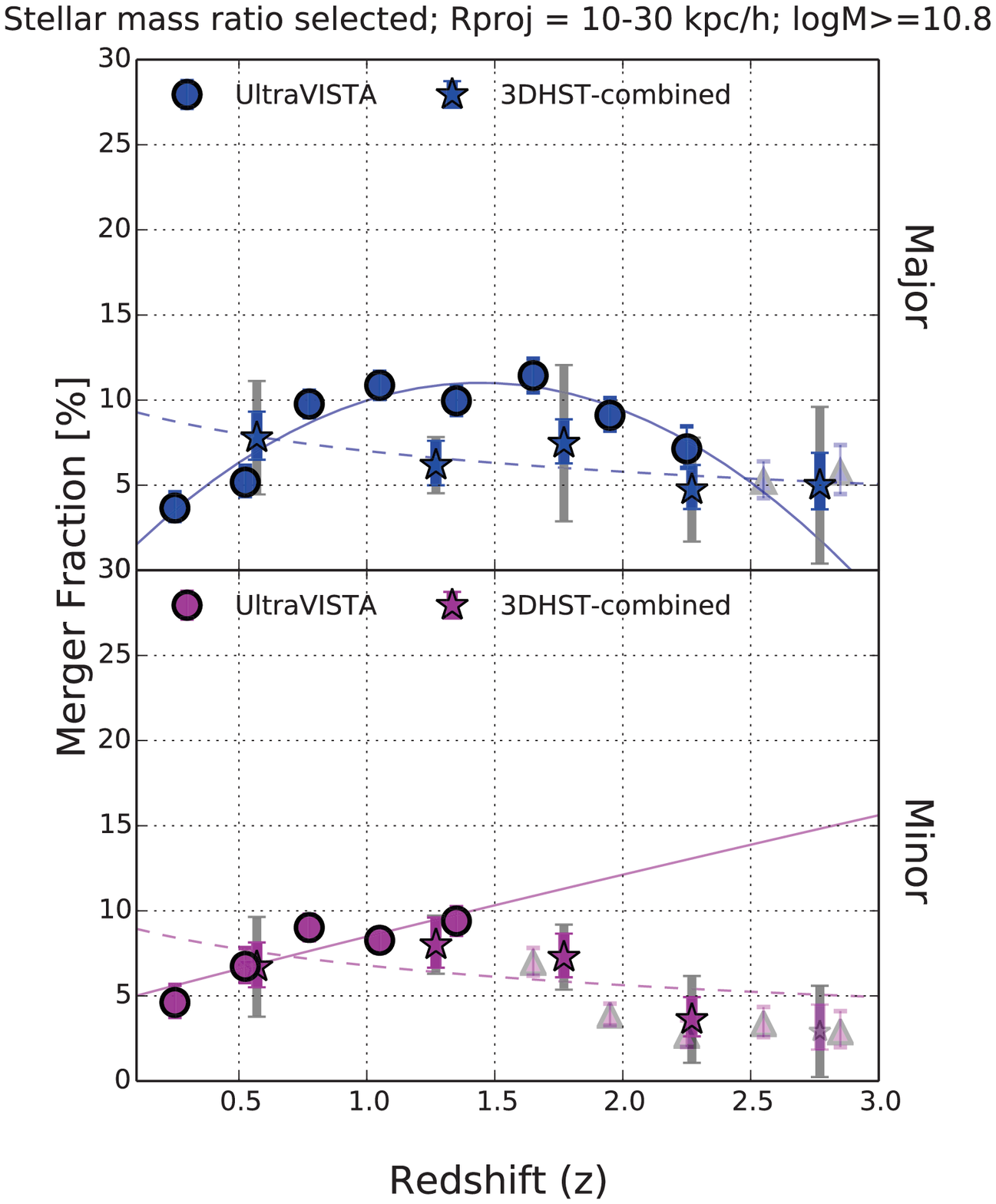} 
	\end{minipage}
	\begin{minipage}[b]{0.47\linewidth}
	\label{fig:pf_fluxratio}
	\centering
	\includegraphics[angle=0,width=\textwidth] {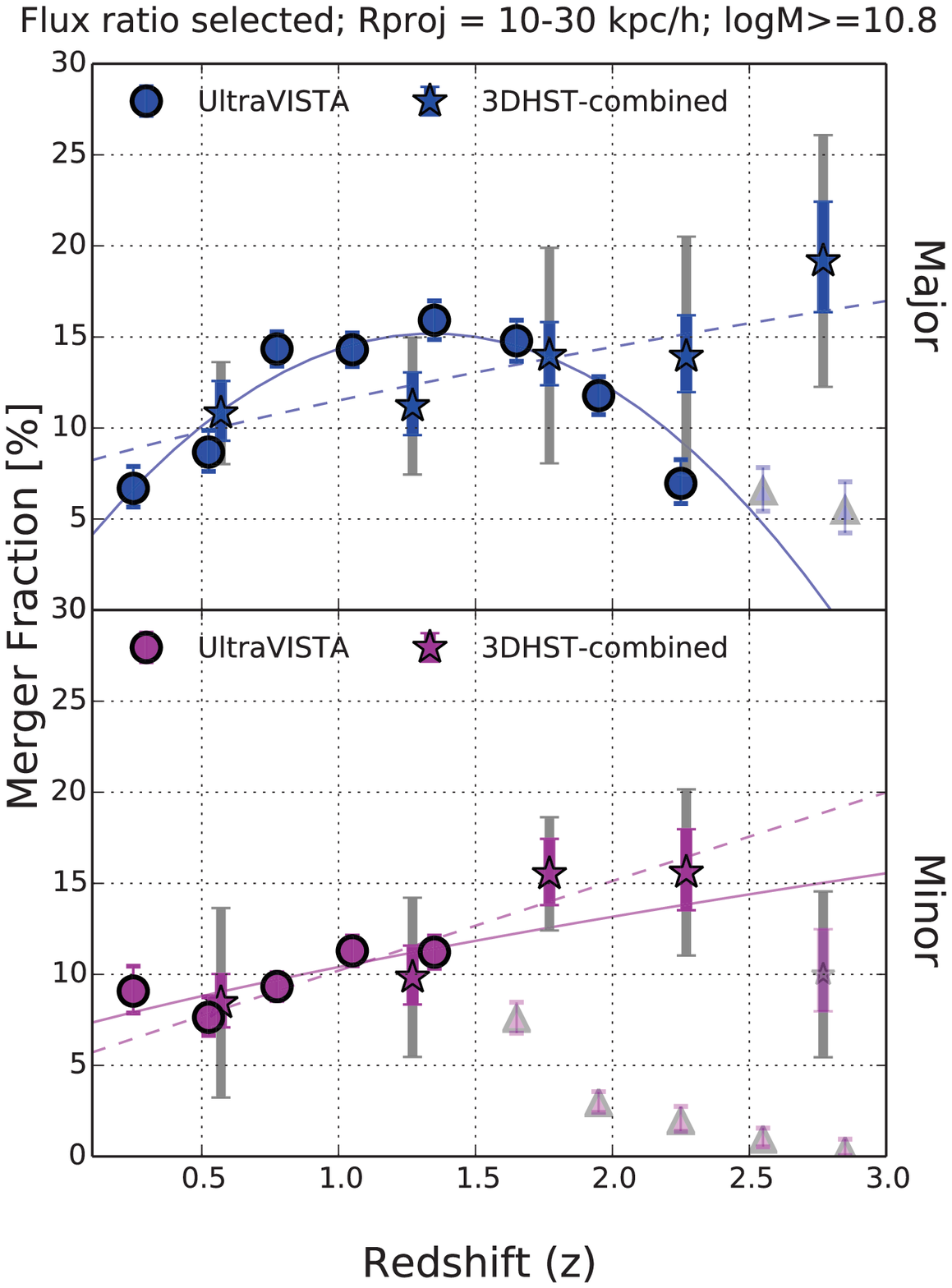}
	\end{minipage}
	\caption{
	The merger fractions of the UltraVISTA (filled circles) and the combined results of the five 3DHST+CANDELS fields (filled stars).
	The left and right panels show the mergers selected by the stellar mass ratio and $H$-band flux ratio respectively, following the definitions in Section~\ref{sec:select_pairs}.
	The top panels show major mergers (stellar mass or flux ratio 1:1 - 4:1) and the bottom panels show minor mergers (stellar mass or flux ratio 4:1 - 10:1) around massive (log$(M_{\star}/M_{\odot}) \geqslant 10.8$) galaxies,
	where the mergers have matching photo-$z$'s and have projected separation between 10 - 30 kpc $h^{-1}$.
	For the 3DHST data points, we combine the pair counts in all five fields,
	and we display the Poisson errors of the combined pair counts of the five fields (colored error bars) as well as the standard deviation of individual measurements from the mean (gray error bars).
	The redshift bins estimated to be incomplete for low surface brightness satellites are marked with semi-transparent upward triangles (UltraVISTA) or small stars (3DHST+CANDELS) following the same color scheme.
	The colored solid and dashed lines are the best-fitting functions to the merger fractions of the UltraVISTA and the 3DHST+CANDELS respectively,
	as presented in Section~\ref{sec:pf_uvista} and Table~\ref{table:bestfit}.
	}
	\label{fig:pf}
\end{figure*}
\section{Method and results}  \label{sec:method}
The relation between the number of observed galaxy pairs ($\mathbb{N}_\mathrm{observed~pairs}$) and the number of ongoing physical galaxy mergers ($\mathbb{N}_\mathrm{physical~mergers}$) can be described as $\mathbb{N}_\mathrm{physical~mergers} = \mathbb{N}_\mathrm{observed~pairs} - \mathbb{N}_\mathrm{projected~pairs} - \mathbb{N}_\mathrm{non-merging~pairs} $.
The quantity $\mathbb{N}_\mathrm{observed~pairs}$ is defined as the number of galaxy pairs observed that satisfy a projected separation and mass (or flux) ratio criteria, 
e.g. pairs fulfiling the first two criteria listed in \ref{sec:select_pairs}.
Among the observed pairs,
some are galaxy pairs of physical proximity,
while some pairs are galaxies projected along a similar line-of-sight.
The line-of-sight projected galaxy pairs can be corrected for using redshift measurements (photometric or spectroscopic) or statistical arguments based on the galaxy mass or luminosity function.
In this work we apply a photo-$z$ criterion as listed in Section~\ref{sec:select_pairs} to correct for projected pairs ($\mathbb{N}_\mathrm{projected~pairs}$).
We have demonstrated in \citet{Man2012} that using the photo-$z$'s to correct for chance alignments yield results consistent with statistical corrections at $z=0-3$.

In this work we do not correct for physical galaxy pairs at matching redshifts that are not energetically bound to merge, 
i.e. we assume $\mathbb{N}_\mathrm{non-merging~pairs}$ = 0.
Cosmological simulations can provide a statistical estimate of $\mathbb{N}_\mathrm{non-merging~pairs}$ to account for the unbound galaxy pairs in cluster environments with high relative velocities.
However, the interpretation may be complicated by the presence of a third neighbor which is not uncommon \citep{Moreno2012, Moreno2013},
or these pairs simply require more time before the eventual coalescence \citep{Kitzbichler2008}.
Galaxy fly-bys may be frequent \citep{Sinha2012} but it remains unexplored how high-speed encounters may impact the mass distribution and light profiles of galaxies.
Even if the cores do not coalesce,
mass from the satellite may still be deposited onto the host galaxy,
and the energy exchange can lead to size growth akin to a ``real'' merger \citep{Laporte2013}.
It is not well understood how $\mathbb{N}_\mathrm{non-merging~pairs}$ evolves with the environment and redshift. 
At higher redshift,
massive galaxies are expected to be less clustered than at the present day,
so the effect is likely more dominant at low redshift.
Future studies of the dynamical properties of galaxy pairs at different redshifts and environments may provide new insights into this effect, 
but for now we do not have enough information to correct for it.
We note that by including non-energetically bound pairs in our selection,
the merger fractions derived in this paper are formally upper limits.
Hereafter we refer to $\mathbb{N}_\mathrm{physical~mergers}$ as $N_{pair}$ for simplicity.

\subsection{Redshift evolution of the merger fraction} \label{sec:pf_uvista}
We define the merger fraction as the fraction of massive galaxies that are merging with a less massive companion,
i.e. $f = {N}_{pair} / N_{massive}$.
The major and minor merger fractions ($f_{major}$ and $f_{minor}$) in redshift bins are listed on Table~\ref{table:pf_massratio} and plotted on Figure~\ref{fig:pf_massratio} (left).
We parameterise the merger fractions within the completeness limits by a power law using least squares fitting.
In the case of $f_{major}$ declining beyond $z\sim1.5$ in UltraVISTA,
the reduced $\chi^{2}$ value for the power law fit exceeds 10 indicating a bad fit so we fit the data points with a quadratic function instead.
We list the best fitting parameters in Table~\ref{table:bestfit}.

Using the stellar mass ratio selection,
we find that $f_{major}$ ($f_{minor}$) increases from $z\sim0.1$ to reach a peak at $z\sim0.8$, 
remains relatively constant to $z\sim 1.7$ ($z\sim 1.4$) and then diminishes towards higher redshift.
A comparison between the merger fractions derived from the ground-based UltraVISTA and the deeper, higher resolution 3DHST+CANDELS reveals very similar $f_{major}$ and $f_{minor}$ in both samples. 
In fact,
$f_{major}$ is slightly lower in 3DHST+CANDELS than in UltraVISTA at $z=1-1.5$.
If we include the pairs without photo-$z$ information (columns 3 and 7 on Table~\ref{table:pf_massratio}) in our merger sample, 
the $f_{major}$ of 3DHST+CANDELS at this redshift bin becomes consistent with the one from UltraVISTA.
This illustrates that space-based data is not required for measuring the galaxy merger fraction.
In fact, ground-based data with a large survey volume such as UltraVISTA provides the optimal dataset,
as the sample is adequately large to measure the redshift dependence of the merger fractions in finer redshift bins.
We elaborate on the uncertainties of merger fraction measurements in Section~\ref{sec:cv}.


\capstartfalse
\begin{deluxetable*}{cccc}
	\tablecolumns{4}
	\tabletypesize{\small} 
	\tablewidth{0pc}
	\tablehead{
	  \colhead{Catalog} &
	  \colhead{Selection} &
	  \colhead{Merger fraction} &
	  \colhead{Merger rate}
	}
	\tablecaption{\textbf{Best fitting functions to merger fractions and rates}}
	\startdata
	\multicolumn{4}{c}{\textbf{Major}} \\ 
UltraVISTA & Stellar mass ratio & $(-5.25\pm0.74)(1+z)^{2}+(25.67\pm3.50)(1+z)+(-20.36\pm3.88)$ & $(0.06\pm0.02)(1+z)^{0.41\pm0.33}$ \\
UltraVISTA & $H_{160}$ flux ratio & $(-7.16\pm1.08)(1+z)^{2}+(33.54\pm5.22)(1+z)+(-24.09\pm5.88)$ & $(0.13\pm0.04)(1+z)^{-0.03\pm0.32}$ \\
3DHST & Stellar mass ratio & $(9.71\pm2.14)(1+z)^{-0.47\pm0.24}$ & $(0.14\pm0.03)(1+z)^{-0.53\pm0.22}$ \\
3DHST & $H_{160}$ flux ratio & $(7.81\pm1.44)(1+z)^{0.56\pm0.19}$ & $(0.11\pm0.02)(1+z)^{0.57\pm0.19}$\\
	\multicolumn{4}{c}{\textbf{Minor}} \\ 
UltraVISTA & Stellar mass ratio & $(4.61\pm0.99)(1+z)^{0.88\pm0.31}$ & $(0.04\pm0.01)(1+z)^{1.07\pm0.26}$ \\ 
UltraVISTA & $H_{160}$ flux ratio & $(6.96\pm1.11)(1+z)^{0.58\pm0.24}$ & $(0.07\pm0.01)(1+z)^{0.55\pm0.27}$\\
3DHST & Stellar mass ratio & $(9.34\pm4.6)(1+z)^{-0.46\pm0.57}$ & $(0.11\pm0.05)(1+z)^{-0.69\pm0.56}$\\
3DHST & $H_{160}$ flux ratio & $(5.21\pm1.51)(1+z)^{0.97\pm0.3}$
 & $(0.06\pm0.01)(1+z)^{0.92\pm0.27}$ \\
	\enddata
	\tablecomments{
	The best fitting functions of the measured merger fractions and rates.
	We quote the power law as long as the reduced $\chi^{2}$ is less than 10,
	and otherwise we use the quadratic function as it proves to be a better fit for the concave shape of the UltraVISTA major merger fractions.
	The parameters are determined by a least square fit to the data points which are complete to low surface brightness satellites.
	We note that the $(1+z)$ dependence are similar for the merger fractions and rates,
	since a constant observability timescale from \citet{Lotz2010a} is applied for the conversion.
	}
	\label{table:bestfit}
\end{deluxetable*}
\capstarttrue
\capstartfalse
\begin{deluxetable*}{ccccccccccc}
	\tablecolumns{11}
	\tablewidth{0pc}
	\tablecaption{ \textbf{Merger fraction: Stellar mass ratio selected}}
	\tablehead{  
	\colhead{} & \colhead{} &  \multicolumn{4}{c}{Major} & \colhead{} & \multicolumn{4}{c}{Minor} \\ 
	\cline{3-6} \cline{8-11} \\ 
	\colhead{Redshift range} &
	\colhead{$N_\mathrm{massive}$} &
	\colhead{$N_\mathrm{match~z} $} &
	\colhead{$N_\mathrm{missing~z} $} &
	\colhead{$N_\mathrm{not~match~z} $} &
	\colhead{$f_\mathrm{major} [\%]$} &
	\colhead{} &
	\colhead{$N_\mathrm{match~z} $} &
	\colhead{$N_\mathrm{missing~z} $} &
	\colhead{$N_\mathrm{not~match~z} $} &
	\colhead{$f_\mathrm{minor} [\%]$}  
	}
	\startdata
	\multicolumn{11}{c}{\textbf{UltraVISTA DR1}} \\ 
$0.1 < z \le 0.4$ & 628 & 23 & 12 & 201 & 3.66$^{+0.93}_{-0.76} $ & & 29 & 6 & 170 & 4.62$^{+1.03}_{-0.85}$ \\ 
$0.4 < z \le 0.65$ & 772 & 40 & 4 & 99 & 5.18$^{+0.96}_{-0.82} $ & & 52 & 3 & 117 & 6.74$^{+1.07}_{-0.93}$ \\ 
$0.65 < z \le 0.9$ & 1618 & 158 & 2 & 170 & 9.77$^{+0.78}_{-0.78} $ & & 146 & 6 & 179 & 9.02$^{+0.75}_{-0.75}$ \\ 
$0.9 < z \le 1.2$ & 1692 & 184 & 6 & 169 & 10.87$^{+0.8}_{-0.8} $ & & 140 & 9 & 140 & 8.27$^{+0.7}_{-0.7}$ \\ 
$1.2 < z \le 1.5$ & 1426 & 142 & 5 & 133 & 9.96$^{+0.84}_{-0.84} $ & & 134 & 10 & 143 & 9.4$^{+0.81}_{-0.81}$ \\ 
$1.5 < z \le 1.8$ & 1163 & 133 & 8 & 99 & 11.44$^{+0.99}_{-0.99} $ & & 81 & 13 & 102 & \dag6.96$^{+0.86}_{-0.77}$ \\ 
$1.8 < z \le 2.1$ & 1087 & 99 & 9 & 97 & 9.11$^{+1.01}_{-0.91} $ & & 42 & 20 & 125 & \dag3.86$^{+0.69}_{-0.59}$ \\ 
$2.1 < z \le 2.4$ & 560 & 40 & 9 & 63 & 7.14$^{+1.32}_{-1.12} $ & & 15 & 4 & 63 & \dag2.68$^{+0.88}_{-0.68}$ \\ 
$2.4 < z \le 2.7$ & 536 & 28 & 13 & 56 & \dag5.22$^{+1.18}_{-0.98} $ & & 18 & 9 & 73 & \dag3.36$^{+0.99}_{-0.78}$ \\ 
$2.7 < z \le 3.0$ & 347 & 20 & 3 & 31 & \dag5.76$^{+1.6}_{-1.28} $ & & 10 & 4 & 46 & \dag2.88$^{+1.23}_{-0.89}$ \\ 

	\multicolumn{11}{c}{\textbf{3DHST COSMOS}} \\ 
$0.1 < z \le 1.0$ & 123 & 10 & 1 & 43 & 8.13$^{+3.46}_{-2.52} $ & & 7 & 1 & 44 & 5.69$^{+3.06}_{-2.09}$ \\ 
$1.0 < z \le 1.5$ & 61 & 3 & 0 & 9 & 4.92$^{+4.79}_{-2.69} $ & & 5 & 0 & 14 & 8.2$^{+5.54}_{-3.54}$ \\ 
$1.5 < z \le 2.0$ & 88 & 5 & 0 & 14 & 5.68$^{+3.84}_{-2.45} $ & & 9 & 1 & 16 & 10.23$^{+4.66}_{-3.34}$ \\ 
$2.0 < z \le 2.5$ & 63 & 6 & 0 & 20 & 9.52$^{+5.68}_{-3.77} $ & & 5 & 1 & 18 & 7.94$^{+5.37}_{-3.43}$ \\ 
$2.5 < z \le 3.0$ & 45 & 0 & 0 & 7 & 0.0$^{+0.0}_{0.0} $ & & 3 & 1 & 7 & \dag6.67$^{+6.49}_{-3.64}$ \\ 

	\multicolumn{11}{c}{\textbf{3DHST GOODS-N}} \\ 
$0.1 < z \le 1.0$ & 84 & 3 & 2 & 22 & 3.57$^{+3.48}_{-1.95} $ & & 6 & 5 & 35 & 7.14$^{+4.26}_{-2.83}$ \\ 
$1.0 < z \le 1.5$ & 72 & 6 & 2 & 13 & 8.33$^{+4.97}_{-3.3} $ & & 7 & 2 & 13 & 9.72$^{+5.23}_{-3.58}$ \\ 
$1.5 < z \le 2.0$ & 57 & 1 & 2 & 7 & 1.75$^{+4.04}_{-1.45} $ & & 3 & 2 & 4 & 5.26$^{+5.12}_{-2.88}$ \\ 
$2.0 < z \le 2.5$ & 65 & 1 & 0 & 6 & 1.54$^{+3.54}_{-1.27} $ & & 1 & 3 & 13 & 1.54$^{+3.54}_{-1.27}$ \\ 
$2.5 < z \le 3.0$ & 37 & 0 & 0 & 6 & 0.0$^{+0.0}_{0.0} $ & & 1 & 0 & 9 & \dag2.7$^{+6.22}_{-2.24}$ \\ 

	\multicolumn{11}{c}{\textbf{3DHST GOODS-S}} \\ 
$0.1 < z \le 1.0$ & 66 & 4 & 0 & 11 & 6.06$^{+4.79}_{-2.89} $ & & 6 & 2 & 25 & 9.09$^{+5.42}_{-3.6}$ \\ 
$1.0 < z \le 1.5$ & 77 & 6 & 1 & 9 & 7.79$^{+4.64}_{-3.08} $ & & 4 & 1 & 11 & 5.19$^{+4.1}_{-2.48}$ \\ 
$1.5 < z \le 2.0$ & 74 & 3 & 0 & 9 & 4.05$^{+3.95}_{-2.22} $ & & 6 & 5 & 12 & 8.11$^{+4.83}_{-3.21}$ \\ 
$2.0 < z \le 2.5$ & 47 & 2 & 0 & 7 & 4.26$^{+5.62}_{-2.77} $ & & 2 & 0 & 8 & 4.26$^{+5.62}_{-2.77}$ \\ 
$2.5 < z \le 3.0$ & 39 & 4 & 0 & 6 & 10.26$^{+8.1}_{-4.9} $ & & 2 & 1 & 7 & \dag5.13$^{+6.77}_{-3.33}$ \\ 

	\multicolumn{11}{c}{\textbf{3DHST AEGIS}} \\ 
$0.1 < z \le 1.0$ & 102 & 8 & 3 & 54 & 7.84$^{+3.86}_{-2.71} $ & & 3 & 5 & 76 & 2.94$^{+2.86}_{-1.61}$ \\ 
$1.0 < z \le 1.5$ & 124 & 7 & 4 & 23 & 5.65$^{+3.04}_{-2.08} $ & & 11 & 10 & 38 & 8.87$^{+3.56}_{-2.63}$ \\ 
$1.5 < z \le 2.0$ & 141 & 19 & 1 & 31 & 13.48$^{+3.85}_{-3.06} $ & & 9 & 5 & 22 & 6.38$^{+2.91}_{-2.08}$ \\ 
$2.0 < z \le 2.5$ & 86 & 5 & 3 & 16 & 5.81$^{+3.93}_{-2.51} $ & & 3 & 1 & 21 & 3.49$^{+3.4}_{-1.91}$ \\ 
$2.5 < z \le 3.0$ & 54 & 4 & 1 & 8 & 7.41$^{+5.85}_{-3.54} $ & & 1 & 3 & 14 & \dag1.85$^{+4.26}_{-1.53}$ \\ 

	\multicolumn{11}{c}{\textbf{3DHST UDS}} \\ 
$0.1 < z \le 1.0$ & 87 & 11 & 6 & 29 & 12.64$^{+5.07}_{-3.75} $ & & 9 & 6 & 19 & 10.34$^{+4.72}_{-3.38}$ \\ 
$1.0 < z \le 1.5$ & 103 & 5 & 2 & 13 & 4.85$^{+3.28}_{-2.1} $ & & 8 & 4 & 17 & 7.77$^{+3.82}_{-2.68}$ \\ 
$1.5 < z \le 2.0$ & 162 & 11 & 2 & 15 & 6.79$^{+2.72}_{-2.01} $ & & 11 & 2 & 38 & 6.79$^{+2.72}_{-2.01}$ \\ 
$2.0 < z \le 2.5$ & 98 & 3 & 0 & 12 & 3.06$^{+2.98}_{-1.67} $ & & 2 & 2 & 10 & 2.04$^{+2.69}_{-1.33}$ \\ 
$2.5 < z \le 3.0$ & 65 & 4 & 0 & 11 & 6.15$^{+4.86}_{-2.94} $ & & 0 & 2 & 16 & \dag0.0$^{+0.0}_{0.0}$ \\ 

	\enddata
	\tablecomments{
	This table presents the number counts of massive galaxies and mergers, 
	as well as the merger fractions in different redshift bins for the UltraVISTA catalog and the five individual fields of the 3DHST+CANDELS catalog.
	The number of massive galaxies is denoted by $N_\mathrm{massive}$.
	The numbers of major (stellar mass ratio 1:1 - 4:1) and minor (stellar mass ratio 4:1 - 10:1) pairs with projected separation $R_{proj} = 10-30$ kpc $h^{-1}$ are further separated according to their photo-$z$ information:
	$N_\mathrm{match~z}$ ($N_\mathrm{not~match~z}$) is the number of pairs with photo-$z$'s (not) matching within their $1\sigma$ uncertainties as described in Section~\ref{sec:select_pairs};
	$N_\mathrm{missing~z}$ is the number of pairs with one or both galaxies not having accurate photo-$z$'s (\texttt{odds} $<0.95$).
	The major and minor merger fractions are calculated as $f = N_\mathrm{match~z} / N_\mathrm{massive}$ in percentages,
	and their uncertainties are propagated from the Poisson errors of $N_\mathrm{match~z}$.
	The $\dagger$ symbols on the merger fractions indicate the redshift bins in which faint, low surface brightness satellites may be incomplete according to Appendix~\ref{sec:sb_limit}.
	}
	\label{table:pf_massratio}
\end{deluxetable*}
\capstarttrue


\subsubsection{Stellar mass ratio or flux ratio?} \label{sec:pf_ratios}

\begin{figure*}[!htp]
	\begin{minipage}[b]{0.5\linewidth}
	\centering
	\includegraphics[angle=0,width=\textwidth]{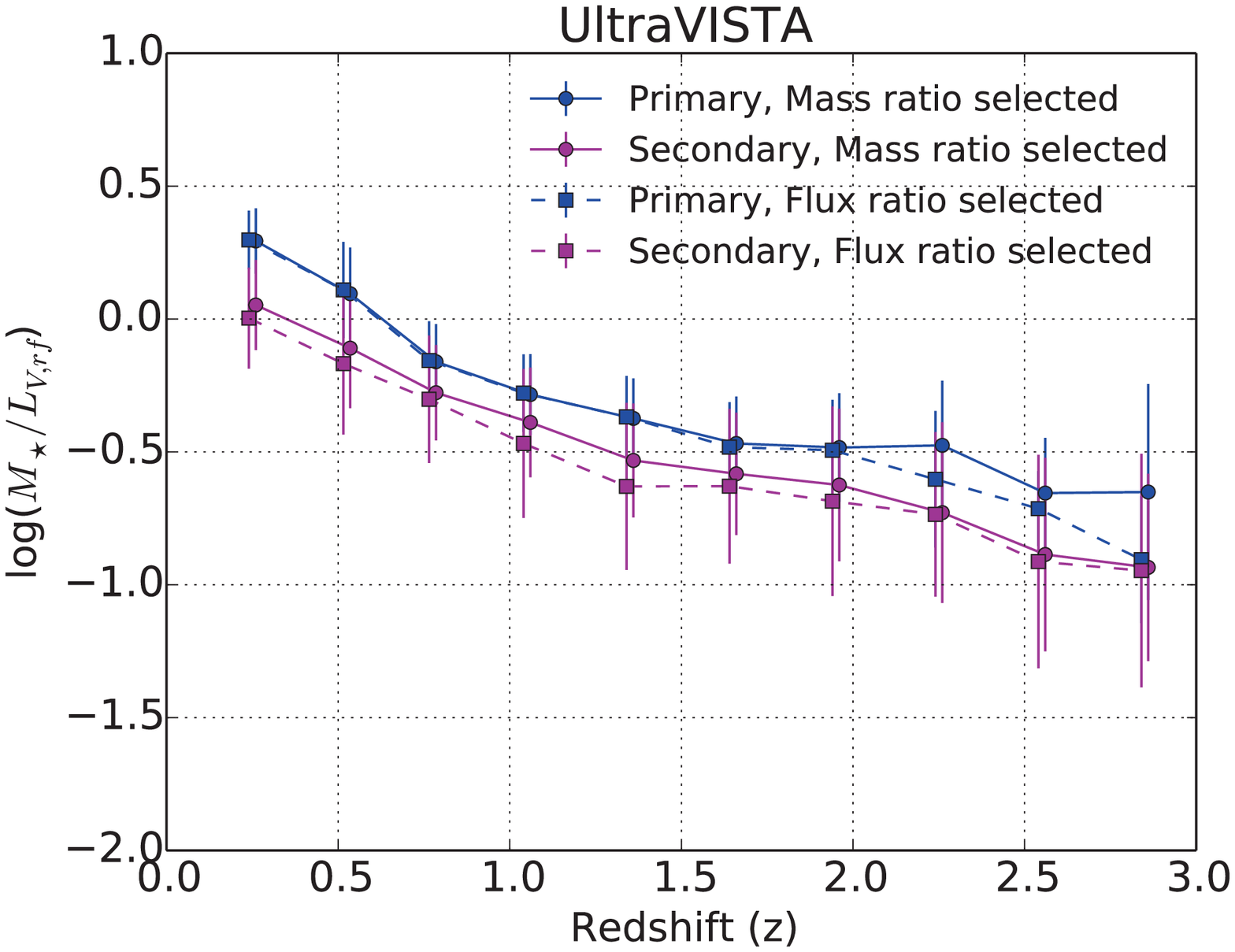}
	\label{fig:MtoL_uvista}
	\end{minipage}
	\begin{minipage}[b]{0.5\linewidth}
	\centering
	\includegraphics[angle=0,width=\textwidth]{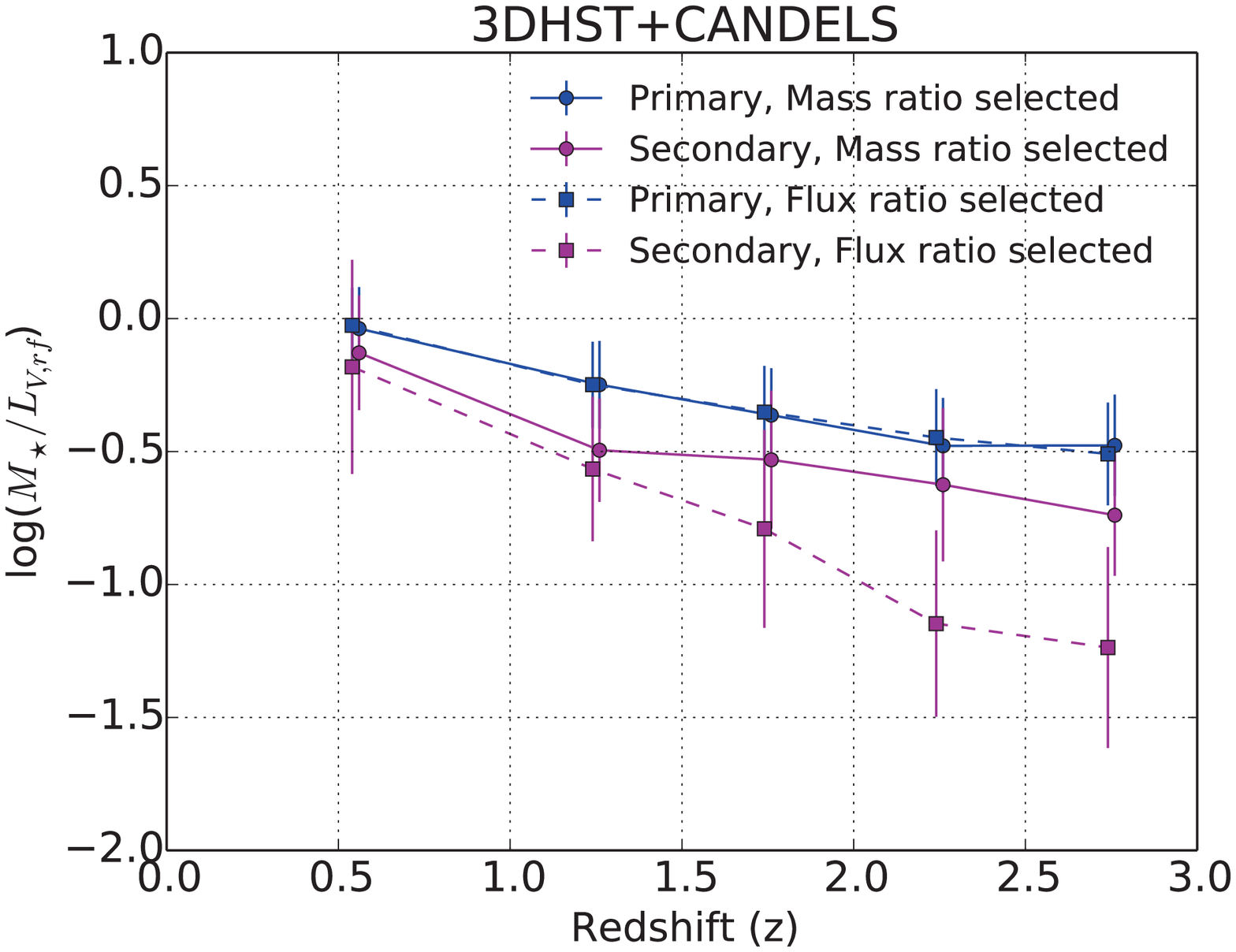}
	\label{fig:MtoL_3dhst}
	\end{minipage}
	\caption{
	We plot the median stellar mass-to-light ratio against redshift for the UltraVISTA and the 3DHST+CANDELS merger samples.
	The stellar mass-to-light ratio ($M_{\star}/L_{V}$) is the stellar mass divided by the luminosity of the rest-frame V-band from InterRest.
	The primary (secondary) galaxies refer to the massive galaxies (satellites),
	and are plotted in blue (magenta).
	We compare the stellar mass ratio (solid) and flux ratio (dotted) selected mergers.
	The error bars show the standard deviation of the $M_{\star}/L_{V}$ in each redshift bin.
	We confirm that for flux ratio selected mergers from 3DHST+CANDELS,
	the $M_{\star}/L_{V}$ of the satellites evolve more steeply than mass ratio selected mergers.
	This supports our finding that the $H$-band flux ratio selection includes satellites with comparable brightness as the massive galaxies, but much lower stellar masses.
	The varying $M_{\star}/L_{V}$ evolution provides the explanation for the discrepancy in the measured merger fractions at $z>2$.
	}
	\label{fig:MtoL_z}
\end{figure*}
\begin{figure}[h!]
	\centering
	\includegraphics[angle=0,width=0.5\textwidth]{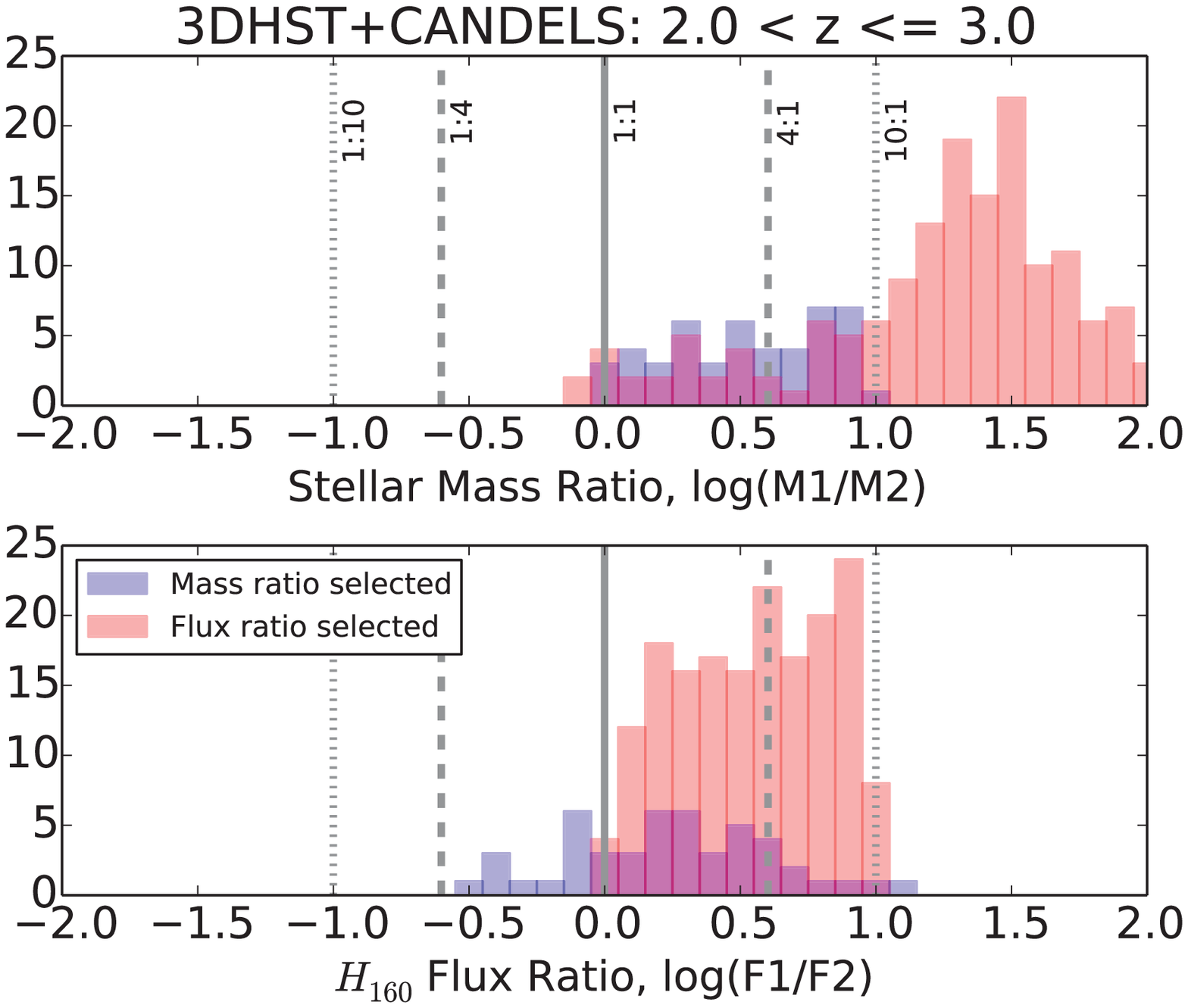}
	\caption{
	These histograms compare the mergers at $2<z\leqslant3$ selected by stellar mass ratio (blue) or $H_{160}$ flux ratio (red) from the 3DHST+CANDELS catalog.
	The ratios are defined such that the mass- (flux-)ratio selected mergers will have ratios of 1 - 10.
	On the top panel we show the histogram of the stellar mass ratios,
	and at bottom the histogram of the $H_{160}$ flux ratios.
	The solid, dashed and dotted gray lines represent the 1:1, 1:4 \& 4:1, 1:10 \& 10:1 ratios respectively.
	From the top panel, 
	we observe that a large excess of flux ratio selected mergers in 3DHST+CANDELS have $H_{160}$ flux ratios between 1 and 10,
	but have stellar mass ratios between 10 and 100.
	This explains the rising merger fractions observed in Figure~\ref{fig:pf_fluxratio} (right) due to bright satellites with log$(M_{\star}/M_{\odot}) < 9.8$ being included in the flux ratio selected sample.
	\label{fig:mr_fr_histogram}
	}
\end{figure}
Merger fraction measurements have led to conflicting conclusions regarding whether it increases with redshift at $z>1.5$ \citep{Bluck2009, Man2012} or not \citep{Williams2011, Newman2012}.
The former studies use the single band flux ratio from \textit{HST} \textit{H}-band imaging to estimate the mass ratio,
rather than full the stellar mass ratio from SED fits used in the latter studies.
We explore the possibility of a systematic effect regarding the ratio used in the merger selection.
We repeat the selection of mergers with the $H_{160}$-band flux ratio instead of using the stellar mass ratio on the same dataset presented in Section~\ref{sec:data}, 
namely the UltraVISTA and 3DHST+CANDELS catalogs.

The results are presented in Figure~\ref{fig:pf_fluxratio} (right).
It is apparent that the combination of using the flux ratio to select mergers and the 3DHST+CANDELS catalog leads to an increasing redshift trend of $f_{major}$ ($f_{minor}$) up to $z=3$ ($z=2.5$) where the catalog is complete for major (minor) satellites.
This is in contrast to the flat or even diminishing evolution found when mergers are selected by the stellar mass ratio (Figure~\ref{fig:pf_massratio}, left),
as well as using flux ratio to select mergers from UltraVISTA (filled circles in Figure~\ref{fig:pf_fluxratio}, right).
Our results are in good agreement with the trends found in literature (see Appendix~\ref{sec:pf_lit} for details of the comparison)
meaning that we are able to reproduce the increasing redshift trend of the merger fraction if mergers are selected by flux ratio.

By comparing the mergers selected in the overlapping area of the UltraVISTA and CANDELS-COSMOS surveys,
we find that the flux-ratio selected satellites at $z>2$ are close to the survey depth of UltraVISTA DR1 ($K \sim 23.4$, \citealt{Muzzin2013}),
and therefore fainter satellites are missed due to low surface brightness.
We interpret the difference between the flux ratio selected merger fraction between the UltraVISTA and the 3DHST-CANDELS samples as being due to the observation limit of the UltraVISTA DR1 data.
This is expected to improve for the forthcoming data release of UltraVISTA in which the survey depth of the four ultra-deep stripes will be $\sim 1$ mag deeper.

In order to explain the difference between the flux and stellar mass ratio selections using the 3DHST+CANDELS catalog,
we compare the stellar mass ratio and flux ratio distribution of the mergers using both selection techniques in Figure~\ref{fig:mr_fr_histogram}.
We display the results for the redshift bin $z=2-3$ where the discrepancy in the merger fraction is most significant between the two selection techniques.
We find that almost all of the stellar mass ratio (1:1-10:1) selected mergers have $H$-band flux ratio in the same range.
On the other hand,
flux ratio selected mergers (1:1-10:1) include mergers with stellar mass ratios in the same range, 
as well as mergers with more extreme stellar mass ratios ($>$10:1).
Among the major \textit{flux ratio} pairs at $z=2-3$ in 3DHST+CANDELS,
only $29\%$ have major \textit{stellar mass} ratios.
The remaining pairs consist of minor \textit{stellar mass} ratio ($19\%$) and mostly very minor \textit{stellar mass} ratio ($52\%$) with $M_{1}/M_{2}>$10:1.
This demonstrates that the observed $H$-band flux is a biased tracer of the stellar mass at $z>2$.
Using the $H$-band flux ratio as a probe for the stellar mass ratio leads to the inclusion of bluer, less massive galaxies as satellites. 
In another words, at $z>2$ most of the satellites are star-forming blue galaxies that are bright in the rest-frame optical B- or V-bands.
We conclude that the flux ratio selection yields a higher merger fraction than mass ratio selection at all redshifts for two reasons:
(1) the observed $H$-band probes bluer rest-frame bands at higher $z$;
(2) lower $M_{\star}/L_{V}$ satellites enter the sample \citep{Bundy2004, Newman2012},
where $M_{\star}/L_{V}$ is the ratio of the stellar mass to the rest-frame $V$-band luminosity.
We illustrate the redshift dependence of $M_{\star}/L_{V}$ in Figure~\ref{fig:MtoL_z}.
There is overall $M_{\star}/L_{V}$ redshift evolution in both the massive galaxies and their satellites,
in which the ratio increases over cosmic time.
Both catalogs show a similar $M_{\star}/L_{V}$ evolution except for the $H$-band flux ratio selected pairs in the CANDELS+3DHST sample,
where the evolution is steeper implying the inclusion of lower $M_{\star}/L_{V}$ at $z>2$ than for the stellar mass ratio selection.
At $2<z\leqslant3$ the observed $H_{160}$-band roughly corresponds to the rest-frame $B$ and $V$ bands.
Our simulations in Appendix~\ref{sec:sb_limit} indicate that we are complete to $z=3 (2.5)$ for major (minor) mergers in 3DHST+CANDELS,
therefore the $M_{\star}/L_{V}$ evolution cannot be explained by observational effects and is intrinsic.
The $M_{\star}/L_{V}$ evolution reflects the higher star formation activity at $z\sim2$ compared to that of the present day \citep[e.g.][]{Lilly1996, Madau1996}.

Having shown that the use of the flux and stellar mass ratio can reproduce the discrepancy in merger fraction in literature,
we proceed to find the ratio that best describes the dynamics and future evolution of the merging galaxies.
Although using the $H$-band flux ratio selection is biased towards star-forming but low stellar mass satellites,
the use of the stellar mass ratio may be biased against gas-rich satellites at $z>1$.
Galaxies appear to be more gas-rich at higher redshift and at lower masses \citep{Erb2006, Mannucci2009, Stewart2009b, Conselice2013}.
Such a dependence implies that the baryon mass ratio is closer to unity than the stellar mass ratio,
since cold gas mass is included into the baryon mass calculation.
The baryon mass of a galaxy is a better probe of its total mass (which also includes dark matter) than the stellar mass alone, as shown in cosmological simulations \citep{Stewart2009b, Hopkins2010b}.
A merger can be \textit{major} or \textit{minor} depending on whether the stellar mass, baryon mass or total mass is considered for the mass ratio \citep{Stewart2009b, Lotz2011}.
Intermediate mass galaxies of log$(M_{\star}/M_{\odot}) \sim 9.8-10.8$ are the satellites to the massive galaxies studied here,
and their molecular gas mass may not be negligible in the total mass budget that governs the dynamics of the galaxies, especially at $z\sim2$.
If the cold gas fraction increases with redshift and decreases with stellar mass as previously claimed \citep{Stewart2009a},
there is a redshift-dependent underestimation if we use the stellar mass to trace the baryon mass.
The correction is likely larger at higher redshift due to the higher gas fraction.
Therefore merging with these gas-rich satellites with stellar mass ratios more extreme than 10:1 may contribute to the star formation budget of the massive galaxies \citep{Conselice2013},
in the form of gas accretion or very minor mergers if characterised by the stellar mass ratio.
We note that gas-rich satellites are not equivalent to gas-rich mergers \citep[e.g.][]{Tadaki2014}, 
which is usually defined as the average gas fraction of both galaxies.
Despite the importance of the gas content in the merger definition as well as its contribution to star formation activity,
direct measurements of the molecular gas mass are so far only available for limited samples of galaxies \citep{Daddi2010, Tacconi2010, Bothwell2013, Tacconi2013},
mostly starbursting sub-millimeter galaxies.
ALMA surveys of large samples of ``normal" star-forming galaxies will shed light on this topic in the future \citep{Scoville2014}.

\subsubsection{Cosmic variance} \label{sec:cv}

\begin{figure*}[!htb]
	\begin{minipage}[b]{0.5\linewidth}
	\centering
	\includegraphics[angle=0,width=\textwidth]{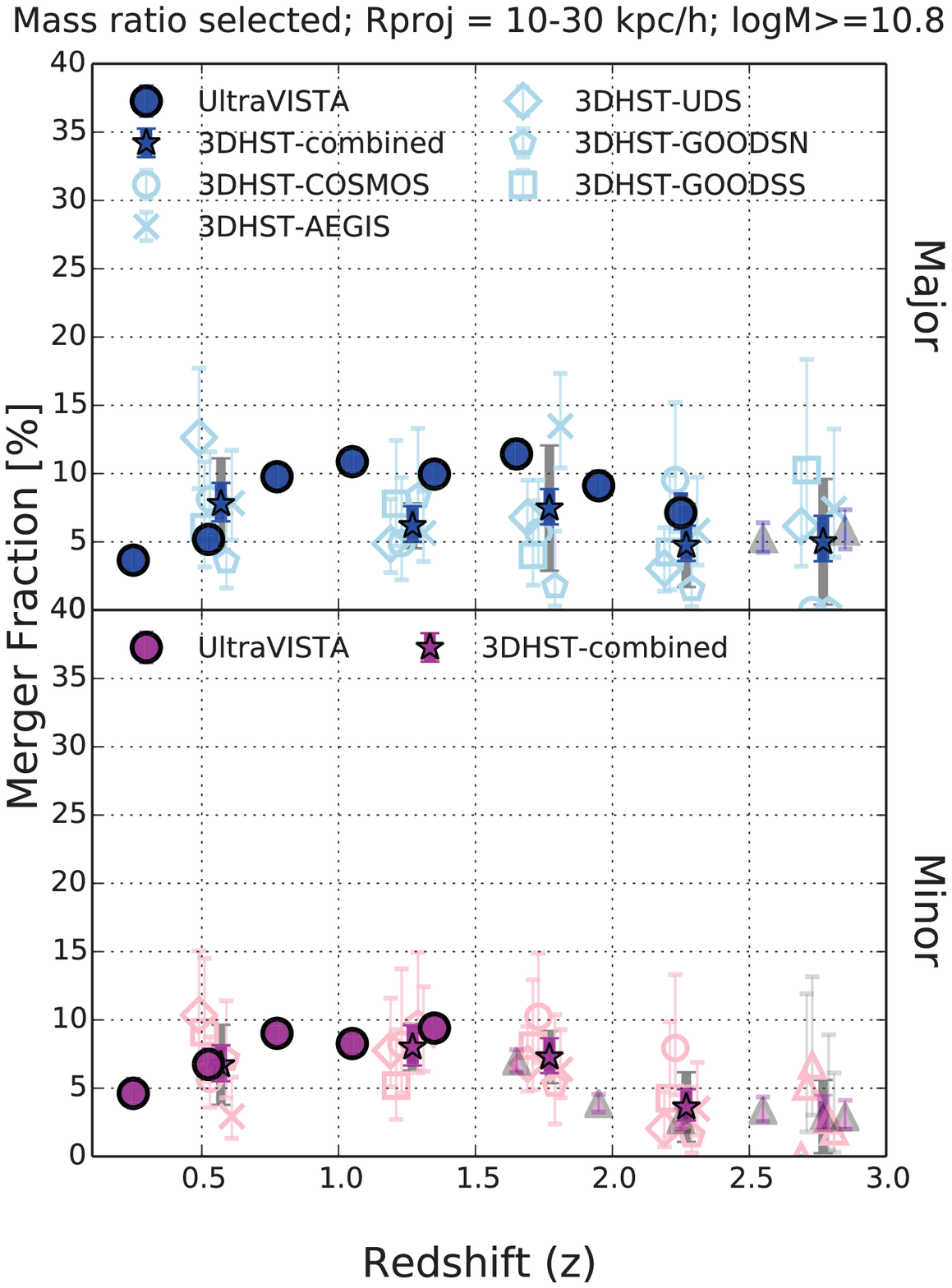}
	\end{minipage}
	\begin{minipage}[b]{0.5\linewidth}
	\centering
	\includegraphics[angle=0,width=\textwidth]{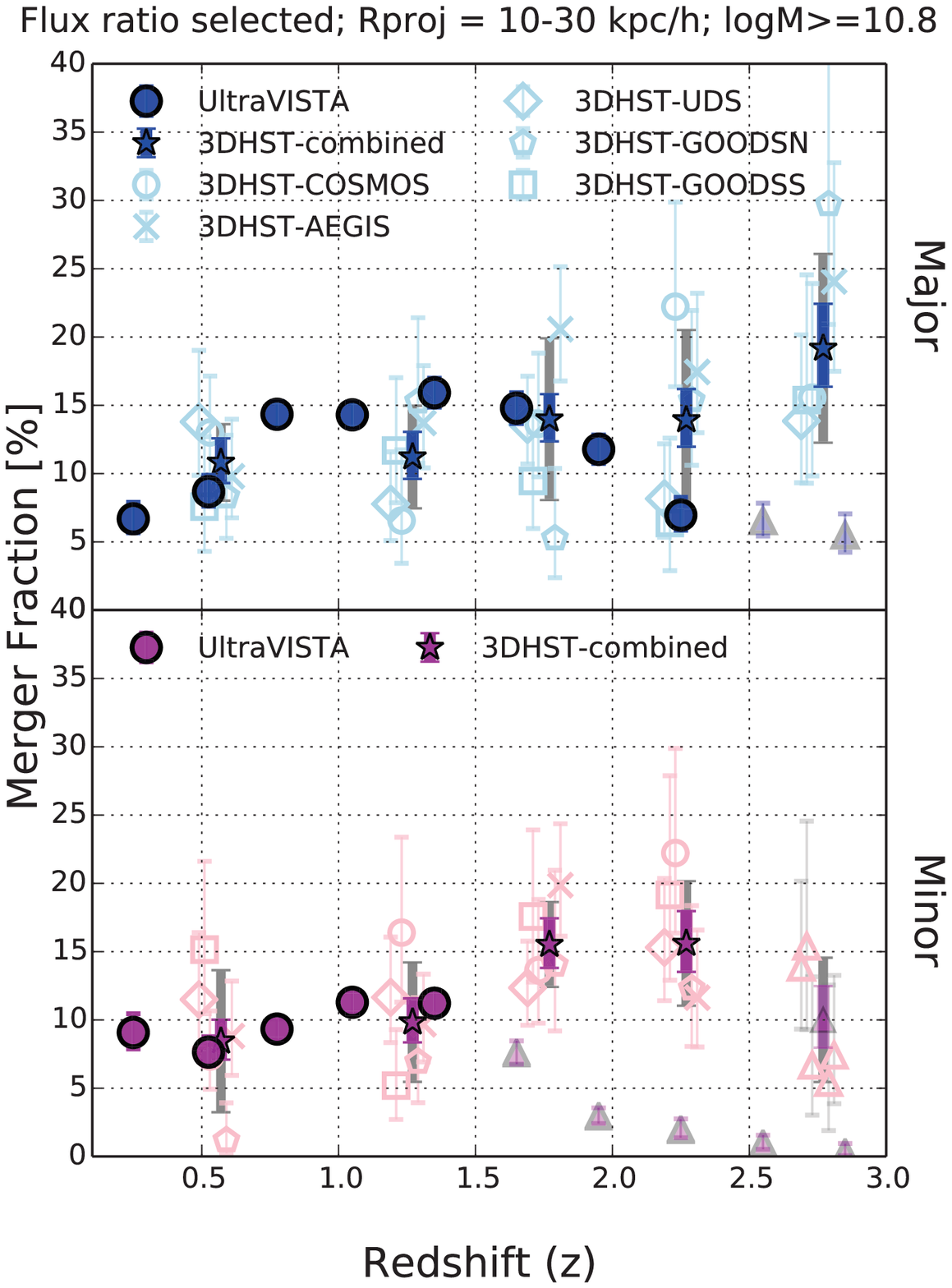}
	\end{minipage}
	\caption{
	The merger fraction measured individually from each of the five 3DHST+CANDELS fields (COSMOS, GOODS-N \& GOODS-S, AEGIS, and UDS) plotted in light blue (major) and pink (minor) with open symbols as indicated in the legend.
 	The combined mean from the five fields are plotted as filled stars.
	The Poisson uncertainties of the combined number of pairs are plotted as the blue / magenta error bars,
	and the standard deviation of the merger fraction of individual fields from the combined mean is shown as gray error bars.
	We can see that cosmic variance is a prominent source of uncertainty for the 3DHST+CANDELS merger fractions.
	The UltraVISTA merger fractions are shown for comparison in filled circles.
	For both catalogs we use triangles to indicate the high redshift regimes in which the catalogs are  estimated to be incomplete for low surface brightness satellites.
	As in the preceding plots, the top panels show major mergers (mass ratio 1:1 - 4:1) and the bottom panels show minor mergers (mass ratio 4:1 - 10:1) around massive (log$(M_{\star}/M_{\odot}) \geqslant 10.8$) galaxies.
	The left plots show the stellar mass ratio selected mergers,
	and the right plots show the $H$-band flux ratio selected mergers.
	The mergers are selected to have overlapping photo-$z$'s and projected separation between 10 - 30 kpc $h^{-1}$ as described in Section~\ref{sec:select_pairs}.
	}
	\label{fig:pf_3dhst}
\end{figure*}
\capstartfalse
\begin{deluxetable*}{ccccccccc} 
	\tablecolumns{9}
	\tablewidth{0pc}
	\tablecaption{Error budget for merger fraction measurements in 3DHST+CANDELS}
	\tablehead{ 
	  \colhead{Redshift range} &
	  \colhead{$f_\mathrm{major} [\%]$} &
	  \colhead{$\sigma_{major, Poisson}$} &
	  \colhead{$\sigma_{major, CV}$} &
	  \colhead{$\sigma_{major, total}$} &
	  \colhead{$f_\mathrm{minor} [\%]$} &
	  \colhead{$\sigma_{minor, Poisson}$} &
	  \colhead{$\sigma_{minor, CV}$} &
	  \colhead{$\sigma_{minor, total}$}
	}
	\startdata
$ 0.1 < z \le 1.0 $ & 7.8 & 0.18 & 0.39 & 0.43 & 6.7 & 0.20 & 0.39 & 0.44 \\ 
$ 1.0 < z \le 1.5 $ & 6.2 & 0.21 & 0.16 & 0.27 & 8.0 & 0.18 & 0.11 & 0.21 \\ 
$ 1.5 < z \le 2.0 $ & 7.5 & 0.17 & 0.59 & 0.61 & 7.3 & 0.18 & 0.19 & 0.26 \\ 
$ 2.0 < z \le 2.5 $ & 4.7 & 0.27 & 0.58 & 0.64 & 3.6 & 0.32 & 0.63 & 0.70 \\ 
$ 2.5 < z \le 3.0 $ & 5.0 & 0.33 & 0.86 & 0.92 & 2.9 & 0.45 & 0.80 & 0.92 \\ 
	\enddata
	\tablecomments{
	A table comparing the dominant sources of uncertainties of the merger fractions for the stellar mass ratio selected mergers in 3DHST+CANDELS fields.
	The fractional error is calculated by the ratio of the error to the merger fraction ($\sigma = \delta f / f$).
	Here we compute the Poisson error of the total pair counts combining the five 3DHST fields ($\delta f_\mathrm{Poisson} = \delta N_\mathrm{pair, Poisson} / N_\mathrm{massive}$).
	The total error is the standard deviation of the merger fraction of each field compared to the combined merger fraction ($\delta f_{total} = \sqrt{\sum_{i=1}^{5}({f_{i} - \overline{f}})^{2}/(5-1)}$).
	The cosmic variance (CV) is calculated by the errors in excess to the expected Poisson errors of the merger fraction in the five fields, 
	i.e. $\sigma_{CV}^{2} = \sigma_{total}^{2} - \sigma_{Poisson}^{2}$.
	The cosmic variance is a dominant source of uncertainty for merger fraction measurements using 3DHST+CANDELS,
	having comparable to or sometimes larger contribution than than the Poisson uncertainty.
	}
	\label{table:3dhst_cv}
\end{deluxetable*}
\capstarttrue

It is apparent from Figure~\ref{fig:pf_3dhst} that a considerable scatter exists for the merger fractions measured in the individual fields of the 3DHST+CANDELS.
The small survey area ($\sim$0.05deg$^{2}$ for each of the five fields) could lead to systematic uncertainties comparable to or larger than the Poisson uncertainties.
We list the fractional errors ($\sigma = \delta f / f$) of the merger fraction measurements of the CANDELS+3DHST sample in Table~\ref{table:3dhst_cv}.
The Poisson uncertainties of the merger fractions are calculated as $\delta f_{Poisson} = \delta N_{pair, Poisson} / N_{massive}$.
We compute the standard deviation of the merger fraction in each field from the combined mean as $\delta f_{total} = \sqrt{\sum_{i=1}^{5}({f_{i} - \overline{f}})^{2}/(5-1)}$,
where $i$ represents the measurement of each of the five fields.
The cosmic variance is simply the observed variance in excess of the Poisson random noise, 
given by $\sigma_{CV}^{2} = \sigma_{total}^{2} - \sigma_{Poisson}^{2}$.
The cosmic variance is a comparable or sometimes larger contributor to the total error budget of the merger fraction measurements than the Poisson uncertainty,
as visualised in Figure~\ref{fig:pf_3dhst}.
More specifically, in the redshift range of $z=1.5-2.0$ the $f_{major}$ measured from AEGIS is $13.5^{+3.9}_{-3.1}\%$, 
whereas the same quantity is measured to be $1.8^{+4.0}_{-1.5}\%$ in GOODS-N.
While each of these quantities are $\sim1.5\sigma$ from the $\overline{f}$ averaged over the five CANDELS fields,
if the individual measurements are taken at face value without including the cosmic variance in the error budget,
the results can differ by a maximum of $\sim7.7\times$ depending on the field used.
Combining the measurements from the five CANDELS fields is crucial to mitigate cosmic variance, also known as the field-to-field variance \citep{Grogin2011}.

The cosmic variance affecting the merger fraction measurements depends primarily on the number densities of the massive galaxies and their satellites, 
as well as the cosmic volume probed, 
as shown by \citet{LSanjuan2014}.
Here we use their parametrisation to estimate the relative cosmic variance for the UltraVISTA and 3DHST+CANDELS samples.
If we assume that the number densities of the massive galaxies and their satellites are not different in UltraVISTA than in the combined five fields of 3DHST+CANDELS,
the cosmic variance has a dependence on the comoving volume as $\sigma_{CV} \propto V_{c}^{-0.48}$.
Since the comoving volume is proportional to the survey area,
and UltraVISTA covers $\sim 6.5 \times$ larger area than the fields of 3DHST+CANDELS combined,
we expect the $\sigma_{CV}$ of UltraVISTA to be $\sim 0.41 \times$ that of 3DHST+CANDELS.
Another prominent error of the merger fraction is the Poisson number count of pairs.
As $\sigma_\mathrm{Poisson}$ is proportional to $1/\sqrt{N_\mathrm{pair}}$,
and again assuming similar number densities of satellites in both fields,
we expect $N_\mathrm{pair} \propto Area$ and therefore the Poisson errors should be $\sim 0.39 \times$ smaller in UltraVISTA than that in 3DHST+CANDELS.
This implies that the total fractional error of merger fraction measured from UltraVISTA to be $56\%$ that of 3DHST+CANDELS.

To summarise, 
we caution against drawing conclusions from merger fraction measurements based on individual CANDELS-sized fields.
The merger fraction measurements from the five 3DHST+CANDELS fields combined are comparable to those from UltraVISTA which covers $\sim6.5\times$ larger area,
albeit with larger Poisson uncertainties and in coarser redshift bins.
We call for including cosmic variance as a systematic uncertainty for pencil beam surveys such as 3DHST+CANDELS for merger fraction measurements \citep{Somerville2004, Moster2011, Xu2012a}.

\subsection{Why are there so few minor mergers?} \label{sec:few_minor}

Minor dry mergers are often invoked as the primary driver of the observed size evolution of quiescent massive galaxies from $z\sim2$ to 0.
Predictions from numerical simulations and virial arguments \citep{Bezanson2009, Naab2009, Laporte2013} suggest that they are more efficient than major dry mergers in puffing up the sizes of quiescent galaxies per unit mass added.
From previous minor merger fraction measurements \citep{Williams2010, Newman2012} and this work (see Section~\ref{sec:merger_rates}) it is inferred that massive galaxies undergo less than one minor merger since $z\sim2$. 
However, if the sole explanation of the observed size evolution is minor merging,
multiple minor mergers are required \citep[e.g.][]{Hilz2012, Oser2012, Hilz2013}.
Here we investigate the possibilities of missing faint satellites to massive galaxies at $z>1.5$.

\subsubsection{Are we missing minor mergers because of observational bias?}

As discussed in Section~\ref{sec:pf_ratios},
we find that neither the major nor minor merger fractions in the CANDELS deep fields are higher than those in the CANDELS wide fields,
although measurements from individual fields are subject to high cosmic variance (see Section~\ref{sec:cv}).
Additionally, the merger fractions from stellar mass ratio selected mergers of UltraVISTA and 3DHST+CANDELS are remarkably consistent (Figure~\ref{fig:pf_massratio}, left),
even in the redshift bins where UltraVISTA is incomplete for low surface brightness galaxies.
Even though the CANDELS \textit{H}-band imaging is $>$3 magnitudes deeper and has $>4\times$ smaller PSF compared to UltraVISTA,
UltraVISTA has the advantage that it probes a redder band (\textit{Ks}) where high redshift galaxies are brighter.

To make a robust claim that we do not miss minor mergers lying just below the surface brightness limits (SB) of our surveys,
we refer to the simulation performed for the completeness limits as introduced in Appendix~\ref{sec:sb_limit}.
In short, we confirm that we do not miss minor mergers up to $z=2.5$ in 3DHST+CANDELS.
We arrive at this conclusion by making the most conservative assumption that the faintest possible satellite is a maximally old, dust-free galaxy of log$(M_{\star}/M_{\odot}) = 9.8$ for a range of light profiles.
The completeness limits hold except for the extreme cases not simulated: 
(1) they have very compact sizes ($R_{e} < 0.39$ kpc) and Sersic index $n>4$ so that they have insufficient contiguous pixels above the detection threshold;
(2) they have very large sizes ($R_{e} > 1.95$ kpc) and low $n<0.5$ so they have low SB;
(3) their dust extinction causes them to be fainter than a dust-free maximally old galaxy.
These size limits are motivated by the scaling relations for quiescent or early-type galaxies \citep{Williams2010, Newman2012, Cassata2013} and simulation assumptions regarding the size of the stellar halo \citep{Hilz2012}.
Unless these intermediate mass galaxies have light profiles very different from the more massive galaxies at similar redshift and similar mass galaxies at lower redshifts,
(1) and (2) are not likely explanations.
The rest-frame optical faintest galaxies at $z>2$ should be quiescent and therefore should be dust-free,
therefore (3) is not a likely explanation either.

From binary merger simulations \citep{Lotz2010a}, 
the observability timescales of major and minor mergers are very short at $R_{proj}<15$ kpc $h^{-1}$ ($<0.1$ Gyr) and therefore we do not expect many close pairs blended by the PSF.
As long as the lower $R_{proj}$ limit for the close pair search is set according to the seeing and SB limit of the data,
the resolution is not expected to cause a bias in the merger fraction.

\subsubsection{What do we expect for the minor merger fraction?} \label{sec:expect_minor_merger}

As lower mass galaxies are more abundant than massive galaxies,
one may expect that minor mergers are more frequent than major mergers from a statistical argument.
Minor mergers are expected to be visible as pairs for longer than major mergers, 
according to dynamical friction timescales arguments and binary simulations \citep{Lotz2010a}.
Therefore one intuitively expects the minor merger fraction and rate to be higher than the major ones.
However, cosmological simulations indicate that the major and minor merger rates are comparable in the stellar mass range probed in this work \citep{Croton2006, Maller2006, Somerville2008, Stewart2009b, Hopkins2010b, Cattaneo2011} due to the stellar mass dependence on the $M_{\star}-M_{halo}$ relation.

\begin{figure}[h!]
	\includegraphics[angle=0,width=0.5\textwidth]{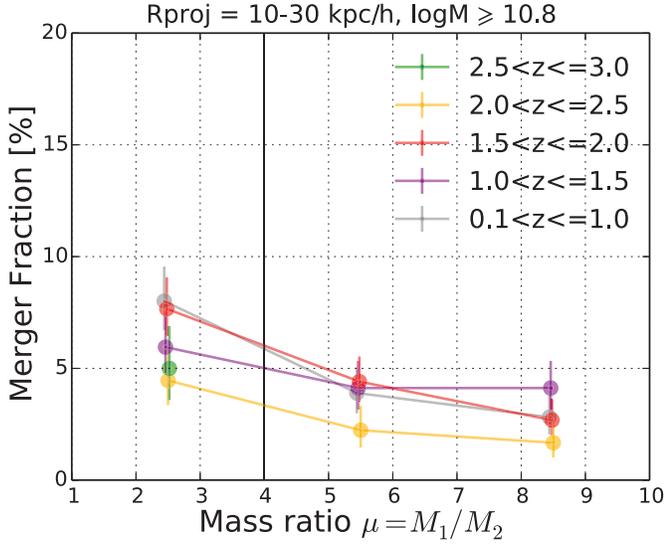}
	\centering
	\caption{
	The dependence of stellar mass ratio selected merger fractions on the stellar mass ratio at different redshifts.
	We note that the distribution of stellar mass ratios is remarkably insensitive to the catalog used (UltraVISTA or 3DHST+CANDELS) and the selection method (stellar mass ratio or $H$-band flux ratio),
	except for a declining tail towards lower stellar mass ratios for the $H$-band flux ratio selection as discussed in Section~\ref{sec:pf_ratios}.
	On this plot we display the stellar mass selected ratio mergers from 3DHST+CANDELS for illustration.
	Only the data points in which they are complete in stellar mass and surface brightness are shown (see Table~\ref{table:completeness_all}).
	The major merger fractions appear to be comparable to the minor merger fractions at all redshifts.
	}
	\label{fig:pf_mu}
\end{figure}

With our large complete sample of mergers, 
we can study the relative fractions of mergers of different stellar mass ratios ($\mu$).
We present our merger fractions in various $\mu$ bins in Figure~\ref{fig:pf_mu}.
The merger fraction decreases as the $\mu$ gets more extreme.
The minor ($4 \leqslant \mu \leqslant 10$) merger fractions are comparable to the major merger ($1 \leqslant \mu \leqslant 4$) fractions at all redshifts.
This is in qualitative agreement with previous observations \citep{LSanjuan2010, Newman2012, Williams2011}.
For our sample of stellar mass ratio selected mergers from both datasets,
the geometric number-weighted mean stellar mass ratio is $<\mu_{n}>~\sim$~4:1 - 5:1 and the mass-weighted mean stellar mass ratio is $<\mu_{m}>~\sim$~3:1 - 4:1.
This is in consistency with various model predictions \citep{Cattaneo2011, Lackner2012, Gabor2012} except \citet{Oser2012},
who find $<\mu_{m}>~\sim$ 5:1 but $<\mu_{n}> \sim$ 16:1.
Their simulation is able to resolve down to 100:1 mergers,
whereas we impose a cut at 10:1 mergers.
We attribute the discrepancy to a higher minor merger rate of their simulated massive galaxies,
as well as our imposed cutoff at $\mu$=10:1.

\subsection{Converting merger fractions to merger rates} \label{sec:merger_rates}

The goal of measuring the galaxy merger fraction is to determine the time integral of the merger rate, defined as the number of mergers ($\mathbb{N}$) that a massive galaxy experiences on average over a time span.
The merger rate can be compared to the observed evolution of the galaxy population, such as in numbers, mass, size, etc., 
so that we can infer if galaxy merging is likely a driver.

Merger rates scale as the number of mergers ($N_{merge, actual}$) occurred during the time span ($\Delta t$) defined by the redshift bin, 
divided by the time span, (Rate $ \propto N_{merge, actual} / \Delta t $).
We measure $\Gamma$ as the number of observed merging galaxies ($N_{merge, obs}$) divided by the observability timescale of mergers ($\tau_{obs}$), 
i.e. Rate $ \propto N_{merge, actual} / \Delta t = N_{merge, obs} / \tau_{obs}$.
The two common definitions of merger rates can be generalized as follows \citep[and references therein]{Lotz2011}: 

(1) The number of merger events per unit time per unit volume ($\Gamma$):
\begin{equation}
	\label{eqt:mergerrate}
	$$ \[ 
	\Gamma (z) [Gyr^{-1} Mpc^{-3}] 
	= \frac{N_{merge,obs}(z) / \tau_\mathrm{obs}}{V_\mathrm{comoving}(z)} 
	= \frac{n_\mathrm{merge}}{\tau_\mathrm{obs} }
	\] $$ 
\end{equation}
where $N_{merge, obs}(z)$ refers to the number of major (or minor) satellites around massive galaxies in that redshift,
$\tau_\mathrm{obs}$ is the average observable timescale for the mergers of the mass ratio range observed to be within $R_{proj}$,
and $V_\mathrm{comoving}$ is the comoving volume projected by the survey area within the concerned redshift interval.

(2) The number of merger events per galaxy per unit time ($\mathbb{R}$) is defined as:
\begin{equation}
	\label{eqt:mergerrate_vol}
	$$ \[ 
	\mathbb{R} (z) [Gyr^{-1}] 
	= \frac{\Gamma (z)}{n_\mathrm{massive}(z)}
	= \frac{n_\mathrm{merge}}{ n_\mathrm{massive} \tau_\mathrm{obs} }
	= \frac{f_\mathrm{merge}}{\tau_\mathrm{obs}}
	\] $$ 
\end{equation}
where $n_\mathrm{massive}$ is the number density of massive galaxies per unit volume.

The number of mergers a massive galaxy undergoes on average ($N_{merger}$) is simply the time integral of the merger rate per galaxy:
\begin{equation}
	\label{eqt:nmerger}
	$$ \[ 
	N_{merger} 
	= \int^{t_{2}}_{t_{1}} \mathbb{R} (z) dt 
	= \int^{z_{2}}_{z_{1}} \frac{\mathbb{R} (z) ~~ t_{H}}{(1+z) ~~ E(z)} dz
	\] $$ 
\end{equation}
where $t_{H}$ is the Hubble time, 
and $E(z) = H(z) / H(0) = [\Omega_{M} (1+z)^{3} + \Omega_{k} (1+z)^2 + \Omega_{\Lambda}]^{1/2}$ \citep{Peebles1993} with the $\Omega$'s denoting the density parameters.

\subsubsection{Merger (observability) timescales} \label{sec:compare_timescales}

Merger rates can be inferred by observing the merger fraction as a function of redshift, 
and then a merging timescale is assumed to convert the fraction to a rate.
The assumed merging timescale either comes from binary merger simulations \citep{Lotz2010a},
cosmological simulations \citep{Kitzbichler2008},
or approximation using the dynamical friction timescale.
Here we briefly discuss the various options and justify the merger timescales used in this work.

The dynamical friction timescale \citep{Bell2006, BKolchin2008, Jiang2008} is a suitable approximation for dark matter halo mergers of large mass ratios (i.e. minor mergers).
However, it remains uncertain whether it can describe mergers with baryons or major mergers in which violent relaxation is the dominant mechanism determining the duration of the merger.

The timescales from binary simulations and cosmological simulations are conceptually distinct.
In binary merger simulations \citep[e.g.][]{Lotz2010a},
two galaxies are set on approaching orbits,
and the observability timescale ($\tau_{obs}$) samples the distribution of pre-coalescence pairs as a function of $R_{proj}$.
The timescale $\tau_{obs}$ is a well-defined quantity which is directly applicable to the merger fraction to rate conversion.
This direct simulation method provides an accurate and comprehensible description of merging for the assumed conditions of relative velocity, gas fraction, morphology, etc.
On the other hand,
merging timescale ($\tau_{merge}$) defined in cosmological simulations \citep{Kitzbichler2008} depends on how the start and end of merging are defined,
for example whether the end is the final coalescence of the two galaxy cores or when most of the mass of the satellite galaxy is deposited onto the massive one.
Another complication is that there are different treatments of mapping stellar masses to the DM halos in cosmological simulations \citep[e.g.][]{Berrier2006, Kitzbichler2008}.
We note that merging timescales for major mergers derived using cosmological simulations are shown to be $\sim 1-2$Gyr longer compared with simulations that include baryons \citep{McCavana2012}.
Most importantly,
$\tau_{obs}$ instead of $\tau_{merge}$ should be used to convert the observed fractions into rates.
Therefore in this work we use the $\tau_{obs}$ from \citet{Lotz2010a}.
The cosmological simulations are useful to weigh the timescales of mergers from binary simulations with different assumptions,
such as gas fraction, orbital parameters,
as discussed in details in \citet{Lotz2011}.
Due to the systematic uncertainties in these assumptions,
as well as random uncertainties due to viewing angles of pairs projected in 2D,
the merging (observability) timescale can only be determined at best to $50\%$ accuracy \citep[and references therein]{Hopkins2010b}.

\subsubsection{Merger rates} \label{sec:compare_merger_rates}
\begin{figure*}[!htb]
	\centering
	\includegraphics[angle=0,width=0.8\textwidth,height=10cm]{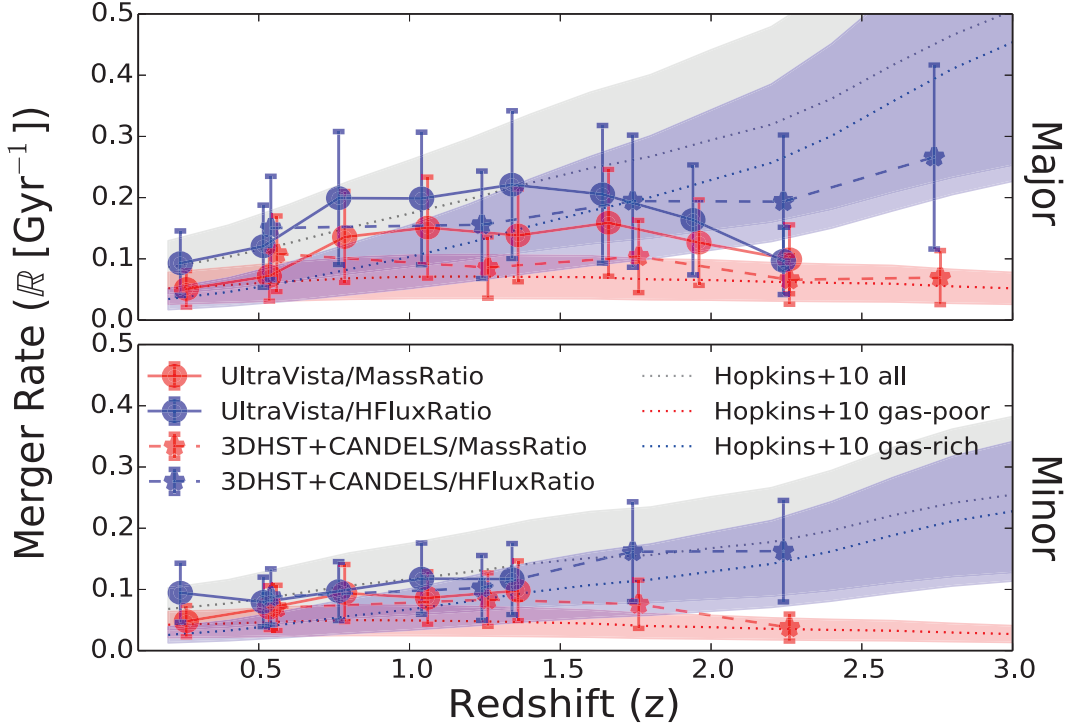}
	\caption{
	The redshift evolution of the major (top) and minor (bottom) merger rates ($\mathbb{R}$) based on our observed merger fractions of UltraVISTA (filled circles, solid lines) and 3DHST+CANDELS (filled stars, dashed lines).
	We compare the merger selections using the stellar mass ratios (red) and $H_{160}$-band flux ratios (blue).
	The merger rates are computed following Equation~\ref{eqt:mergerrate_vol} using the 10-30 kpc $h^{-1}$ close pairs and the observability timescale of \citet{Lotz2010a}.
	The data points are only plotted in the redshift range in which we are complete in detecting the faintest possible satellites.
	We overplot the predicted galaxy merger rates from the simulation of \citet{Hopkins2010a} for comparison.
	The predicted merger rates are plotted as dotted lines with the shades indicating the $50\%$ uncertainties,
	where gray represents mergers of all gas fractions ($f_{gas}$),
	and red and blue represents gas-poor ($f_{gas} = 0-20\%$) and gas-rich ($f_{gas} = 20-100\%$) merger rates.
	}
	\label{fig:merger_rates}
\end{figure*}

\begin{figure*}[!htb]
	\begin{minipage}[b]{0.5\linewidth}
	\centering
	\includegraphics[angle=0,width=\textwidth]{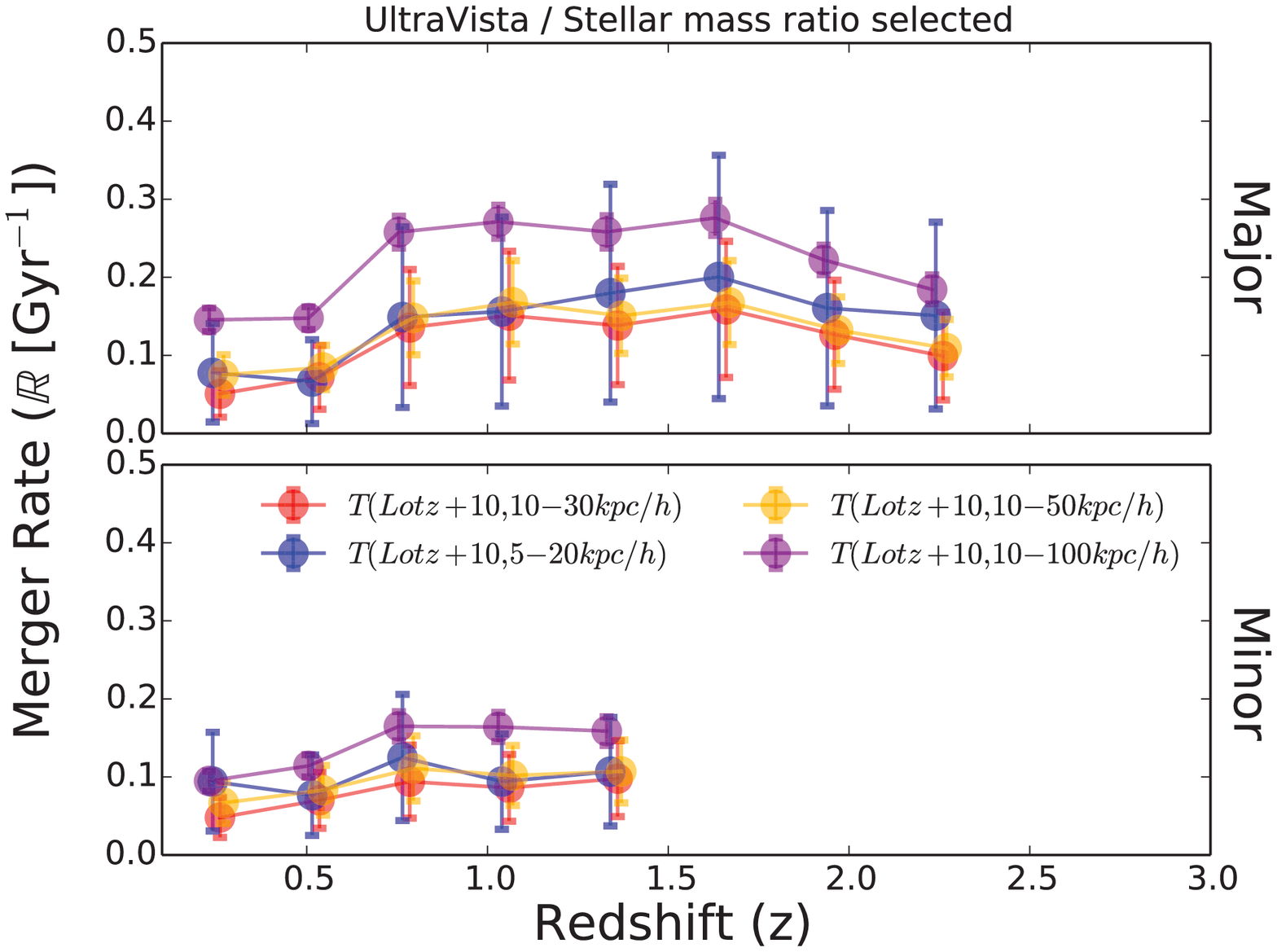}
	\end{minipage}
	\begin{minipage}[b]{0.5\linewidth}
	\centering
	\includegraphics[angle=0,width=\textwidth]{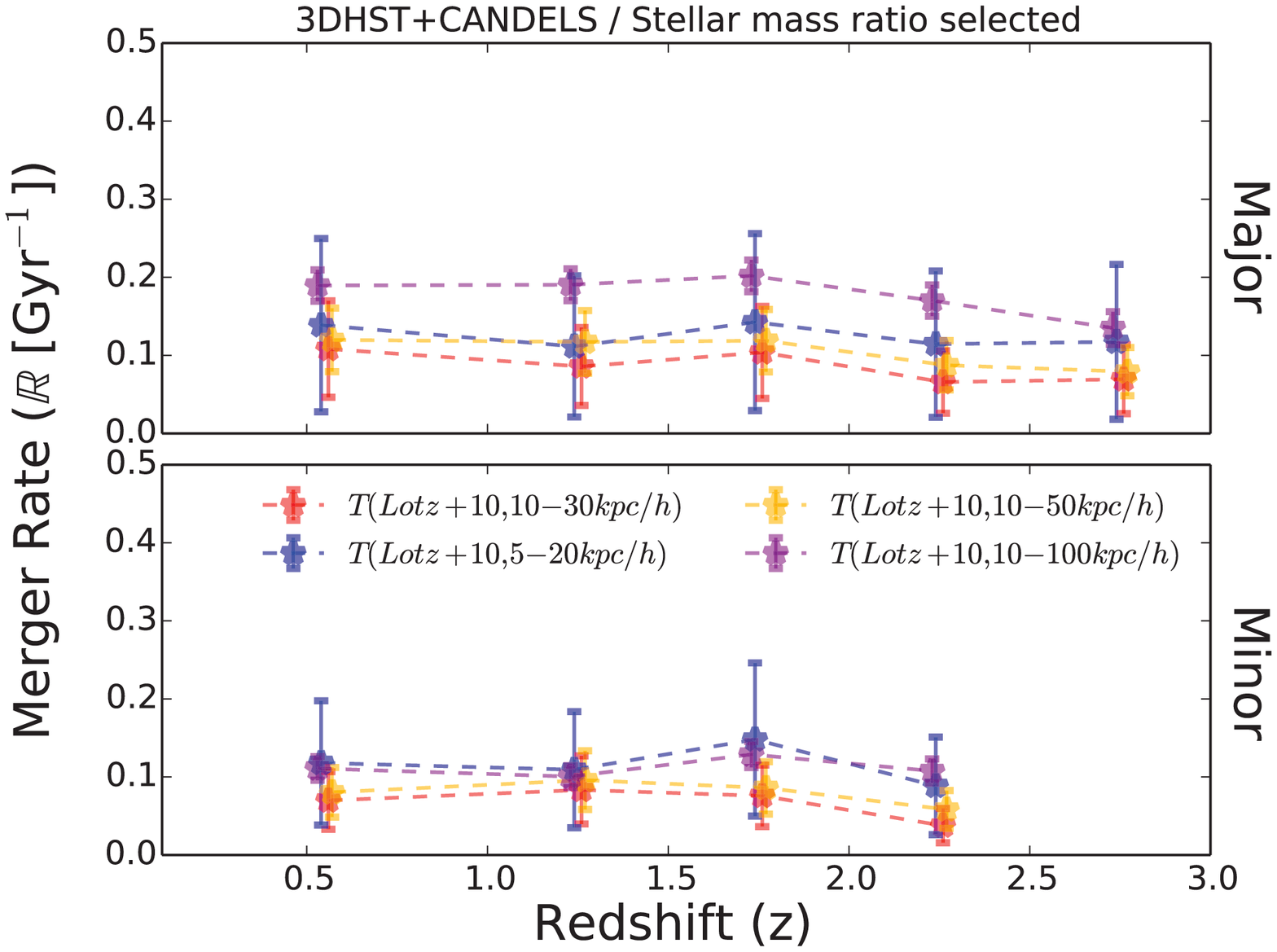}
	\end{minipage}
	\caption{
	Similar to Figure~\ref{fig:merger_rates},
	these figures present the major (top) and minor (bottom) merger rate evolution.
	We compare the merger rates inferred from galaxy pairs of different $R_{proj}$ bins: 
	10-30 kpc $h^{-1}$ (red, default), 5-20 kpc $h^{-1}$ (blue), 10-50 kpc $h^{-1}$ (yellow) and 10-100 kpc $h^{-1}$ (purple).
	The left and right figures show the results of the stellar mass ratio selected mergers in UltraVISTA and 3DHST+CANDELS respectively.
	Only the redshift bins in which the satellites are complete are plotted.
	We demonstrate that the inferred merger rates are consistent within the uncertainties as long as the appropriate observability timescale ($\tau_{obs}$) is applied for the $R_{proj}$ range \citep{Lotz2010a}.
	We note that the merger rates appear to be systematically higher for the widest $R_{proj}$ bin (10-100 kpc $h^{-1}$) compared to the others,
	which we interpret as being due to the wide pairs at $\sim 50-100$ kpc $h^{-1}$ probing the large-scale environment in which pairs at similar redshifts may not necessarily merge within the $\tau_{obs}$ predicted in binary merger simulations.
	}
	\label{fig:merger_rates_diffT}
\end{figure*}

The merger rates derived using Equations~\ref{eqt:mergerrate_vol} and \ref{eqt:nmerger} normalised to timescales of 1 Gyr are shown in Table~\ref{table:merger_rates_normalized}.
We plot the inferred merger rates on Figure~\ref{fig:merger_rates}.
As expected from the merger fractions,
we find the merger rates from UltraVISTA are consistent with those from 3DHST+CANDELS within the completeness range,
and that the flux ratio selection method gives an increasing trend while the stellar mass ratio selection method gives a flat or diminishing trend for the 3DHST+CANDELS catalog.
We list the best fitting parameters for the observed merger rates to a power law in Table~\ref{table:bestfit} for easy comparison to literature.
As the merger rate uncertainties are considerably larger than the measured merger fractions due to the $50\%$ uncertainty in $\tau_{obs}$,
the redshift dependence is weaker and we therefore deem a quadratic fit which has one more degree of freedom than the power law unnecessary.
We show the integrated number of major and minor mergers in Table~\ref{table:Nmerger} for the two catalogs and selection methods.

We find that at $z>2$ the observed merger rates using the stellar mass ratio selection are lower than predicted from the semi-analytical models (SAMs) of \citet{Hopkins2010a, Hopkins2010b} as shown in Figure~\ref{fig:merger_rates},
but are consistent with the gas-poor merger rate ($f_{gas} < 20\%$, where the gas fraction $f_{gas}$ is defined as the ratio of the total gas mass to the total baryon mass of the merging galaxies).
In general the SAMs predict that the galaxy merger rates increase monotonically with redshift.
Our measurements using the $H$-band flux ratio selection show an increasing trend similar to the gas-rich merger rate of \citet{Hopkins2010b} ($f_{gas} \geqslant 20\%$),
even though the $H$-band flux is not a direct tracer of cold gas mass or star formation rate.
This lends support to our claim in Section~\ref{sec:pf_ratios} that using the stellar mass ratio as a probe for the baryon mass ratio may be subject to a bias against gas-rich mergers at $z>2$,
an epoch at which cold gas fraction is non-negligible especially for intermediate mass galaxies \citep{Stewart2009b, Hopkins2010a}.

We also compare the merger rates inferred from the merger fractions of various $R_{proj}$ bins in Figure~\ref{fig:merger_rates_diffT}.
We only show results for the stellar mass ratio selection, 
but the following conclusions also hold for the $H$-band flux ratio selection.
We find that the merger rates are consistent for different $R_{proj}$ bins once the suitable observability timescales from \citet{Lotz2010a} are applied.
On average, the merger rates derived from mergers with $R_{proj}$ = 10-100 kpc $h^{-1}$\footnote{An upper limit of $R_{proj} < 100$ kpc $h^{-1}$ is still small compared to the typical photo-$z$ uncertainty. The typical photo-$z$ error at $z=0-4$ is $\delta z/(1+z) = 0.026$ \citep{Muzzin2013}, corresponding to 84 Mpc/h at $z=1.5$. 
Therefore we do not expect the photo-$z$ uncertainty to constrain widely separated pairs.} are up to $40\%$ higher than for smaller $R_{proj}$ bins, 
although still consistent within the large uncertainties due to the 50\% uncertainty in the merger observability timescale.
This implies that there are more widely separated mergers ($R_{proj} = 50-100$ kpc $h^{-1}$) than expected from the timescales of binary merger simulations.
Possible explanations could be:
 (1) the large scale environment of galaxies are probed at separations of $>50$ kpc $h^{-1}$, 
 therefore we may include galaxies in the same over-densities that are not bound to merge;
 (2) the merging observability timescales for wide pairs may be systematically longer than the assumed tilted polar orbit for close pairs,
 e.g. relative velocities of merging pairs are higher than assumed in the binary simulations (typically $<500$km s$^{-1}$) which may be true in over-densities, 
 or if the merger orbit is more like a circular orbit the merging timescale can be up to $>40\%$ longer \citep{Lotz2010a}.
We note that the discrepancy is larger at lower redshift, 
hinting that the effect could be related to large-scale structure formation.
Cosmological simulations may provide estimates of these effects.
Although we do not use the timescale of \citet{Kitzbichler2008} for galaxy merger fraction measurements for the reasons explained in Section~\ref{sec:compare_timescales},
for comparison we note that using it leads to lower merger rates than those derived using the shorter timescales of \citet{Lotz2010a} as expected from the inverse scaling between timescale and rate.
\capstartfalse
\begin{deluxetable*}{cccccc} 
	\tablecolumns{6}
	\tablewidth{0pc}
	\tablecaption{Merger number densities and rates}
	\tablehead{ 
		\colhead{}    &  
		\multicolumn{2}{c}{Major merger} &   
		\colhead{}   &  
		\multicolumn{2}{c}{Minor merger} \\ 
		\cline{2-3} \cline{5-6} \\ 
		\colhead{Redshift range} & 
		\colhead{$n_{merge} = \Gamma \times \tau_{obs}$[Gyr]}  &
		\colhead{$\Delta N_{merger} \times \tau_{obs}$}  &
		\colhead{} & 
		\colhead{$n_{merge} = \Gamma \times \tau_{obs}$[Gyr]}  &
		\colhead{$\Delta N_{merger} \times \tau_{obs}$} \\
		\colhead{} & 
		\colhead{[ $ \times 10^{-3}$~Mpc$^{-3} h^{3}$] }  &
		\colhead{} &
		\colhead{} &
		\colhead{[ $ \times 10^{-3}$~Mpc$^{-3} h^{3}$] }  &
		\colhead{} 
		}

	\startdata
	\multicolumn{6}{c}{\textbf{UltraVISTA}} \\ 
$0.1 < z \le 0.4$ & 0.111$\pm0.026$ & 0.109$\pm0.025$ & &0.14$\pm0.029$ & 0.138$\pm0.028$ \\ 
$0.4 < z \le 0.65$ & 0.076$\pm0.013$ & 0.09$\pm0.015$ & &0.098$\pm0.015$ & 0.117$\pm0.017$ \\ 
$0.65 < z \le 0.9$ & 0.182$\pm0.015$ & 0.125$\pm0.01$ & &0.169$\pm0.014$ & 0.116$\pm0.01$ \\ 
$0.9 < z \le 1.2$ & 0.131$\pm0.01$ & 0.123$\pm0.009$ & &0.1$\pm0.008$ & 0.094$\pm0.008$ \\ 
$1.2 < z \le 1.5$ & 0.084$\pm0.007$ & 0.083$\pm0.007$ & &0.08$\pm0.007$ & 0.079$\pm0.007$ \\ 
$1.5 < z \le 1.8$ & 0.072$\pm0.006$ & 0.072$\pm0.006$ & &0.044$\pm0.005$ & 0.044$\pm0.005$ \\ 
$1.8 < z \le 2.1$ & 0.051$\pm0.005$ & 0.045$\pm0.005$ & &0.021$\pm0.004$ & 0.019$\pm0.003$ \\ 
$2.1 < z \le 2.4$ & 0.02$\pm0.003$ & 0.028$\pm0.005$ & &0.008$\pm0.002$ & 0.01$\pm0.003$ \\ 
$2.4 < z \le 2.7$ & 0.014$\pm0.003$ & 0.016$\pm0.003$ & &0.009$\pm0.002$ & 0.011$\pm0.003$ \\ 
$2.7 < z \le 3.0$ & 0.01$\pm0.003$ & 0.015$\pm0.004$ & &0.005$\pm0.002$ & 0.007$\pm0.003$ \\ 
\\
	\multicolumn{6}{c}{\textbf{3DHST+CANDELS}} \\ 
$0.1 < z \le 1.0$ & 0.115$\pm0.021$ & 0.5$\pm0.09$ & &0.099$\pm0.019$ & 0.43$\pm0.084$ \\ 
$1.0 < z \le 1.5$ & 0.066$\pm0.014$ & 0.096$\pm0.02$ & &0.086$\pm0.016$ & 0.124$\pm0.023$ \\ 
$1.5 < z \le 2.0$ & 0.08$\pm0.014$ & 0.073$\pm0.013$ & &0.078$\pm0.014$ & 0.071$\pm0.012$ \\ 
$2.0 < z \le 2.5$ & 0.033$\pm0.009$ & 0.031$\pm0.008$ & &0.025$\pm0.008$ & 0.024$\pm0.008$ \\ 
$2.5 < z \le 3.0$ & 0.024$\pm0.008$ & 0.023$\pm0.008$ & &0.014$\pm0.006$ & 0.013$\pm0.006$ \\ 

	\enddata
	\tablecomments{
		This table lists the number density of the stellar mass ratio selected major and minor mergers using the UltraVISTA and 3DHST+CANDELS catalogs.
		The number density $n_{merger}$ is related to $\Gamma$ (number of mergers per unit volume per unit time) and the merger observability timescale $\tau_{obs}$ by $\Gamma (z) = n_{merger} (z) / \tau_{obs}$ as explained in Equation~\ref{eqt:mergerrate}.
		Therefore $n_{merger}$ can be interpreted as the merger rate $\Gamma$ normalized to $\tau_{obs}$ of 1 Gyr.
		The average number of merger experienced in the redshift bin is $\Delta N_{merger}$, 
		calculated by integrating the volume-averaged merger rate $\mathbb{R}$ over the elapsed time ($\Delta N_{merger}
		= \int^{t_{2}}_{t_{1}} \mathbb{R} (z) dt 
		= \int^{t_{2}}_{t_{1}} f_{merge} dt / \tau_{obs} $ 
		if constant $\tau_{obs}$ is assumed ) as described in Equation~\ref{eqt:nmerger}.
		}
	\label{table:merger_rates_normalized}
\end{deluxetable*}
\capstarttrue

\capstartfalse
\begin{deluxetable*}{cccccc} 
	\tablecolumns{6}
	\tablewidth{0pc}
	\tablecaption{The average number of mergers experienced by a massive galaxy during $z=0.1-2.5$}
	\tablehead{  
		\colhead{$R_{proj}$}
		& \multicolumn{2}{c}{Stellar mass ratio selected} 
		&& \multicolumn{2}{c}{$H$-band flux ratio selected} \\
		\cline{2-3} \cline{5-6} \\ 
		& \colhead{Major merger} 
		& \colhead{Minor merger}
		&& \colhead{Major merger}
		& \colhead{Minor merger}  }

	\startdata
	\multicolumn{6}{c}{\textbf{UltraVISTA}} \\ 
10-30 kpc $h^{-1}$ & $0.9\pm0.2$ & $0.7\pm0.1$ && $1.4\pm0.3$ & $0.9\pm0.2$ \\
10-100 kpc $h^{-1}$ & $1.9\pm0.1$ & $1.2\pm0.1$ & & $3.2\pm0.1$ & $1.8\pm0.1$ \\
\\
	\multicolumn{6}{c}{\textbf{3DHST+CANDELS}} \\ 
10-30 kpc $h^{-1}$ & $1.0\pm0.4$ & $0.7\pm0.2$ && $1.5\pm0.6$ & $1.0\pm0.3$ \\
10-100 kpc $h^{-1}$ & $1.8\pm0.1$ & $1.1\pm0.1$ && $2.8\pm0.2$ & $2.2\pm0.2$ \\

	\enddata
	\tablecomments{
	The average number of mergers ($N_{merger}$) experienced by a massive galaxy.
	We calculate $N_{merger}$ by measuring the galaxy merger fraction using galaxy mergers within the stated $R_{proj}$ bins,
	converting the merger fraction into merger rate using a observability timescale for that $R_{proj}$ bin \citep{Lotz2010a} and integrating over cosmic time.
	The $N_{merger}$ derived from all the $R_{proj}$ bins are consistent within the uncertainties except for the widest bin of $R_{proj} = 10-100$ kpc $h^{-1}$, 
	therefore we show the $N_{merger}$ for 10-30 kpc $h^{-1}$ as default and omit the other two bins (5-20 kpc $h^{-1}$ and 10-50 kpc $h^{-1}$) that give consistent results.
	}
	\label{table:Nmerger}
\end{deluxetable*}
\capstarttrue

\subsection{Merger-driven stellar mass accretion rate} \label{sec:mass_growth}

\begin{figure*}[Htb]
	\begin{minipage}[b]{0.48\linewidth}
	\centering
	\includegraphics[angle=0,width=\textwidth]{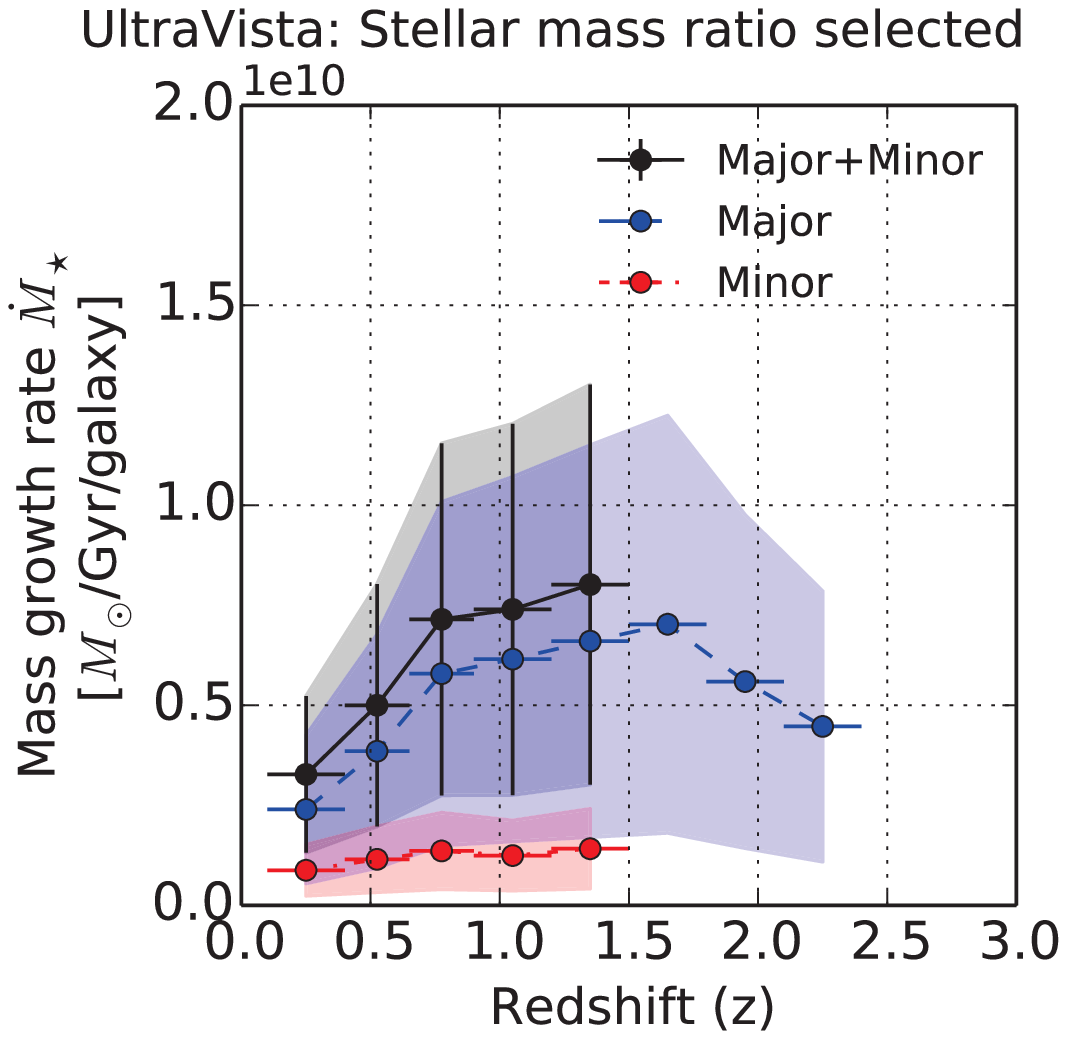}
	\end{minipage}
	\begin{minipage}[b]{0.52\linewidth}
	\includegraphics[angle=0,width=\textwidth]{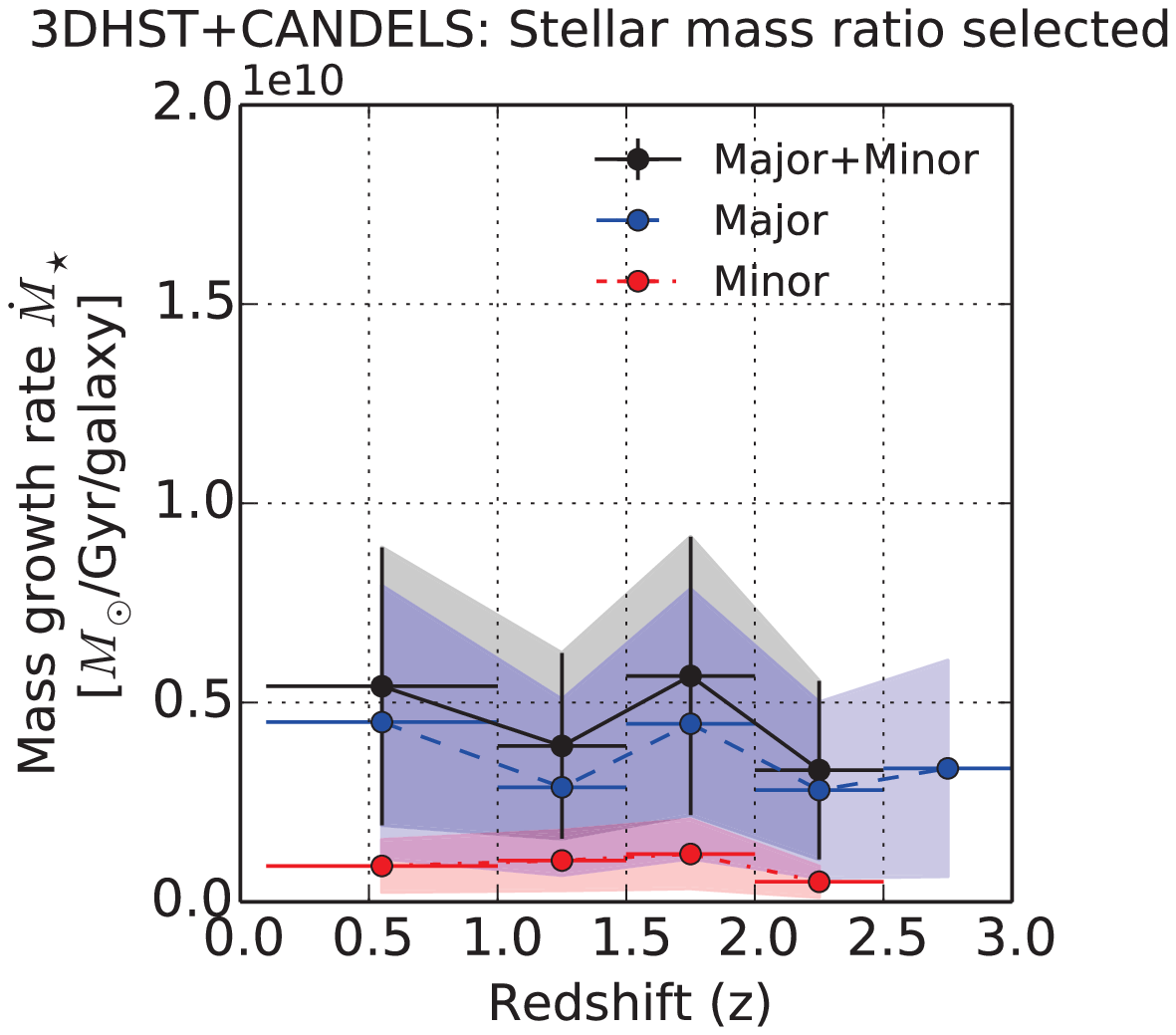}
	\end{minipage}
	\caption{
	The stellar mass growth rate ($\dot{M}_{\star}$) due to the accretion of existing stars via merging is computed as $\dot{M}_{\star}$ [$M_{\odot}$ / Gyr / galaxy] = $\bar{M_{1}} \mathbb{R} / \bar{\mu}$,
	where $\bar{M_{1}}$ is the median stellar mass of the massive galaxies,
	$\mathbb{R}$ is the major (minor) merger rate and $\bar{\mu}$ is the median stellar mass ratio of the major (minor) mergers.
	The results from both the UltraVISTA (left) and 3DHST+CANDELS (right) surveys are shown.
	The blue, red, and black circles denote the stellar mass accretion rate via major, minor merging, and the two combined.
	The shaded regions indicate the uncertainties propagated from the merger rates and stellar masses.
	Only the redshift bins which are complete are plotted.
	We observe that major merging is the primary mechanism for driving the stellar mass accretion of massive galaxies.
	Following the trend of the merger fractions,
	the stellar mass growth rate rises from $z\sim0.1$ to $z\sim0.8$ and remains relatively flat thereafter,
	as seen on the results from UltraVISTA (left).
	There are insufficient galaxies to probe any redshift trend below $z\sim1$ in 3DHST+CANDELS.
	}
	\label{fig:mass_growth}
\end{figure*}

We compute the merger-driven stellar mass accretion rate as
$\dot{M}_{\star}$ [$M_{\odot}$ / Gyr / galaxy] = $\bar{M_{1}} \mathbb{R} / \bar{\mu}$,
where $\bar{M_{1}}$ is the median stellar mass of the massive galaxies,
$\mathbb{R}$ is the major (minor) merger rate,
and $\bar{\mu}$ is the median stellar mass ratio of the major (minor) mergers.
All these quantities are redshift dependent so we are able to calculate the merger-driven stellar mass growth as a function of time.

There is controversy regarding whether merging triggers significant star formation episodes compared to isolated galaxies \citep[e.g.][]{Patton2011, Xu2012b, Yuan2012, Lanz2013, Patton2013, Lackner2014, Puech2014}.
\citet{Gallazzi2014} study the evolution of the age-, mass-metallicity relation of massive galaxies since $z\sim0.7$ to $z\sim0$,
and report that neither new star formation nor chemical enrichment is needed for the evolution of massive quiescent galaxies.
Additionally, we do not have measurements of the gas fraction of our merger sample.
Therefore we note that our analysis only accounts for the accretion of existing stars and ignores stars formed during mergers,
setting the lower limit on the merger contribution to the stellar mass growth.

We show the stellar mass accretion rate as a function of redshift in Figure~\ref{fig:mass_growth}.
For the average massive galaxy of log$(M_{\star}/M_{\odot}) \geqslant 10.8$,
we find that major (minor) merging leads to an average stellar mass growth of 
and $ 4.0 (0.9) \times10^{10}M_{\odot}$ during $z=0.1-2.5$.
This amounts to a total of $4.9\times10^{10}M_{\odot}$ being accreted via 1:1 - 10:1 mergers,
implying that the average $10^{11}M_{\odot}$ galaxies increase their stellar masses by at least $\sim50\%$ through accreting existing stars from satellite galaxies from $z=2.5$~to~0.1.

Our results are in agreement with similar observations for bright central galaxies in galaxy clusters \citep{Lidman2013} and field galaxies \citep{Bundy2004, Ferreras2013} up to $z\sim1$,
showing that major merging plays a significant role in the mass assembly of massive galaxies (and therefore its number density evolution) independent of the environment.
Our stellar mass accretion rates are also consistent with simulation predictions \citep{Stewart2009a, Cattaneo2011, Lackner2012, Laporte2013} with the exception of \citet{Oser2010}.
\citet{Oser2010} follow the history of simulated massive galaxies and find that 
by $z=0$, 80\% of the stars in massive galaxies are formed at $z=3-4$ ex-situ of the original halo at $z=7$,
and are accreted at $z<2$ with an average rate of $\sim 17 M_{\odot}$/yr.
Their average mass accretion rate stays relatively flat at $z>2$ and decreases at lower redshift,
which is qualitatively similar to our observed trends but on average $\geqslant 2\times$ higher,
as seen in Fig.~\ref{fig:mass_growth}.
As we discussed in Section~\ref{sec:expect_minor_merger},
this is explained by the higher minor merger rates in their simulations compared to the observations of this works and others.
We emphasise that the stellar mass accretion rate presented here does not include new stars formed due to merger-triggered star formation episodes, 
and therefore represents a lower limit of the true merger-driven stellar mass growth rate (see also the discussion in Section~\ref{sec:pf_ratios}).

\subsection{Maximum merger-driven size and velocity dispersion evolution} \label{sec:size_evolution}

\begin{figure*}[Htb]
	\begin{minipage}[b]{0.46\linewidth}
	\centering
	\includegraphics[angle=0,width=\textwidth]{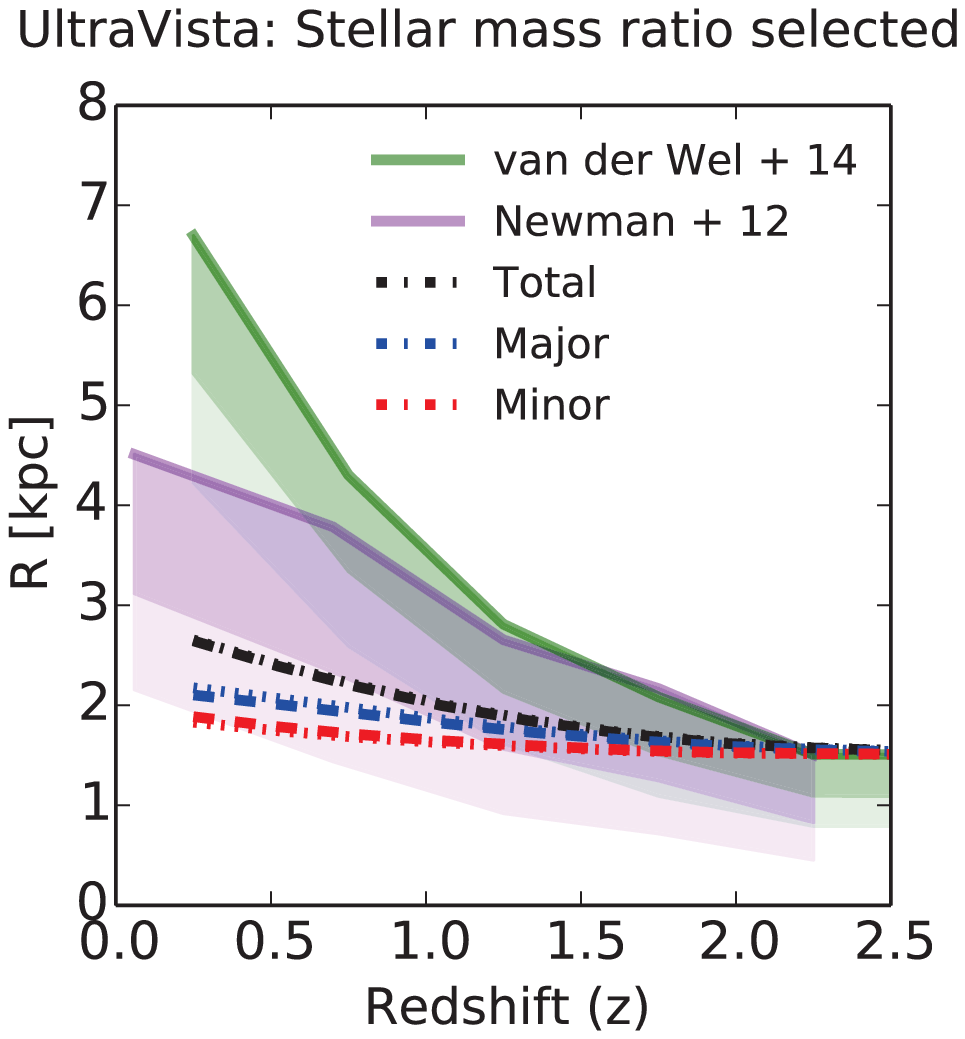}
	\end{minipage}
	\begin{minipage}[b]{0.55\linewidth}
	\centering
	\includegraphics[angle=0,width=\textwidth]{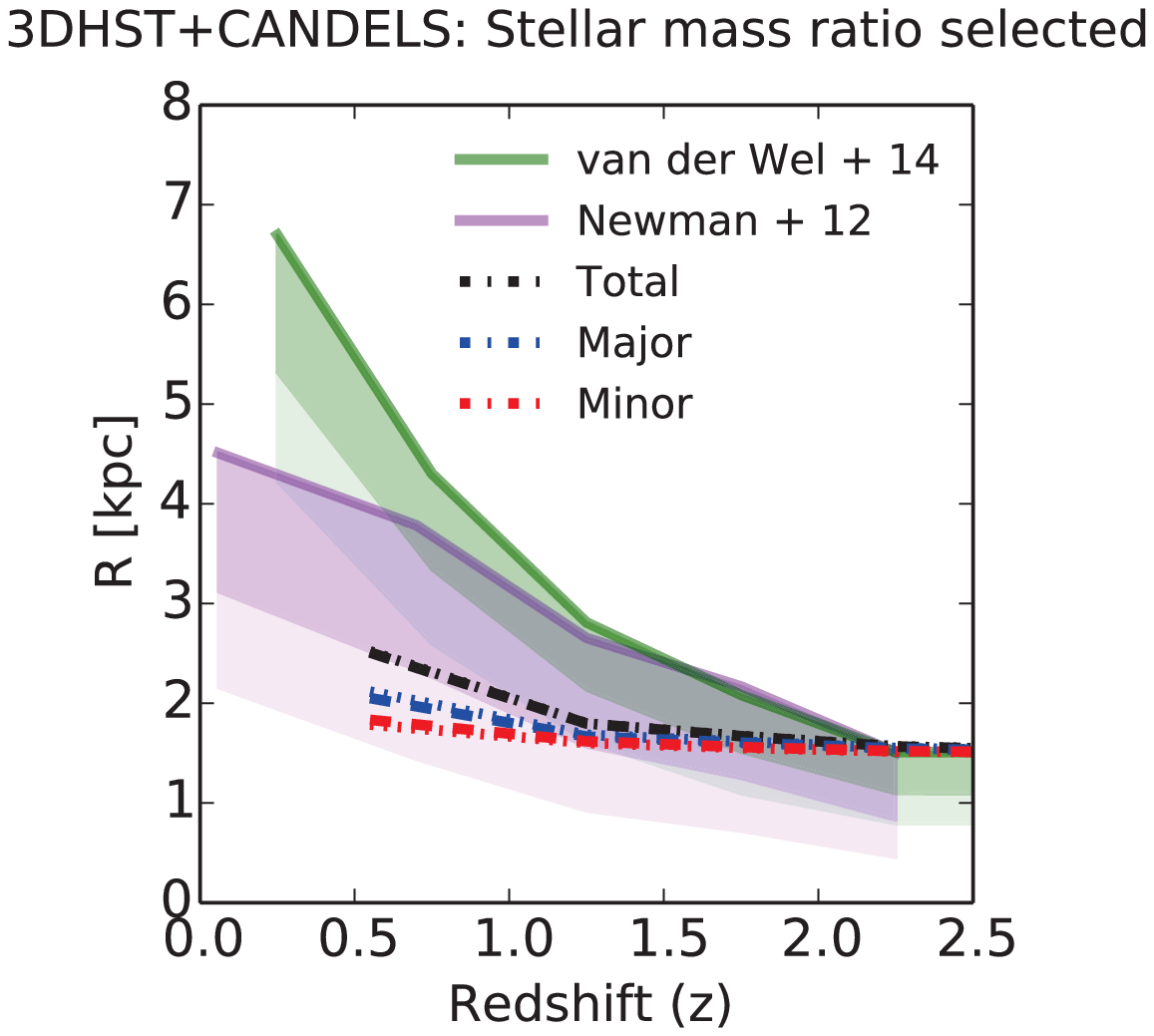}
	\end{minipage}
	\caption{
	The size evolution inferred from the merger-driven stellar mass accretion rate.
	The results from both the UltraVISTA (left) and 3DHST+CANDELS (right) surveys are shown.
	The blue and red lines show the predicted size evolution from major and minor mergers respectively,
	and the black lines are the combined contribution from major and minor mergers.
	We have used two size evolution models:
	the dotted lines represent the virial argument \citep{Naab2009} and the dashed lines represent the model of \citealt{Hilz2013}.
	The size evolutions predicted from both models nearly overlap each other,
	illustrating that the virial theorem is an adequate approximation.
	The models are normalised to a $M_{\star}=10^{11}M_{\odot}$ galaxy of 1.5 kpc at $z=2.5$.
	We compare the predicted merger-driven size evolution with observations:
	the green and orchid lines denote the observed size evolution of early-type / quiescent massive galaxies measured by \citealt{vanderWel2014} and \citealt{Newman2012} respectively.
	The lower $1\sigma$ and $2\sigma$ scatters of the relations are shown by the darker/lighter shades.
	As the merger fractions are consistent between the UltraVISTA and the 3DHST+CANDELS surveys,
	the predicted size evolution are very similar as expected.
	We claim that major and minor merging can increase the sizes of massive QGs by a factor of $\sim2$ at most from $z\sim 2.5$ to 0.
	While this amount of merging is insufficient to explain the observed evolution of the average sizes of massive QGs,
	it is enough to bring the sizes to $1\sigma$ below the mean sizes if the size scales with redshift as $R\propto (1+z)^{-1}$ (\citealt{Newman2012}, see also \citealt{Toft2009, Williams2010, Toft2012, Krogager2013}).
	}
	\label{fig:size_evolution}
\end{figure*}

Dry merging provides a channel to increase the sizes of compact ($\sim1$~kpc) massive quiescent galaxies (QGs) at $z>2$ by a few factors to $z\sim0$ \citep[e.g.][]{Bezanson2009, Naab2009, Oser2012, Hilz2012, Hilz2013},
as discussed in Section~\ref{sec:few_minor}.
We use our measured stellar mass accretion rate to infer an upper limit on the size evolution due to ``dry'' dissipationless merging. %
Since QGs are expected to remain quiescent for the build-up of the red sequence,
and the dissipation from gas in merging galaxies can reduce the efficiency of puffing up sizes of galaxies,
for this exercise we make the simplistic assumption that all observed mergers are dissipationless.
The aim of the test is to investigate to what extend the observed frequency of galaxy merging can explain the size evolution of QGs.
We find that the merger fractions of massive galaxies and the quiescent subset are consistent within their uncertainties,
therefore we simply use the merger fractions of the overall massive galaxy population in the following analysis.

The virial theorem and more sophisticated merger simulations have been used to predict the size evolution due to dry merging.
The size evolution can be parameterised as $R \propto M^{\alpha}$,
where $\alpha \sim 1$ for major merging and $\alpha \sim 2$ for minor merging predicted using the virial theorem \citep{Bezanson2009, Naab2009},
or alternatively $\alpha \sim 0.91$ for major merging and $\alpha \sim 2.3$ for minor merging according to the simulations of \citet{Hilz2012, Hilz2013}.
The high value of $\alpha$ for minor merging in \citet{Hilz2013} implies that it is very efficient in increasing the sizes of galaxies,
and likely represents an upper limit due to the high dark matter content and extended stellar haloes of the satellites assumed in their simulation.
For each redshift bin, 
we multiply the average stellar mass accretion rate (see Section~\ref{sec:mass_growth}) with the time elapsed in the redshift bin to get the stellar mass accreted,
and scale the predicted size growth to the stellar mass accretion using the $\alpha$ values as discussed above.
The maximum merger-driven size growth using both catalogs are plotted in Figure~\ref{fig:size_evolution}.
We observe that the total amount of merging can only increase the size of massive QGs by a factor of two,
from 1.5 kpc at $z=2.5$ to $\sim 3$ kpc at $z\sim0$.
This result is insensitive to the size growth model used,
meaning that the virial theorem provides a good approximation of the size evolution due to dissipationless merging.

The observed size evolution of massive QGs (or early-type galaxies) has been presented in various works.
On Figure~\ref{fig:size_evolution} we compare our predicted merger-driven size evolution to two recent measurements using CANDELS.
\citet{Newman2012} report an average size growth of $\sim3.5$ from $z=2.5$ to 0,
with a redshift dependence of $R\propto(1+z)^{-1.0}$,
consistent with previous works including \citet{Toft2009, Williams2010, Toft2012} and \citet{Krogager2013}.
On the other hand,
\citet{vanderWel2014} report a consistent but slightly stronger size growth of $\sim 5$ times in the same redshift range,
with a redshift dependence of $R\propto(1+z)^{-1.3}$,
similar to the finding of \citet{Cassata2013}.
Both works report the scatter of the stellar-mass size relation to be consistent with being constant.
The difference of the observed size evolution may be due to the stellar mass threshold,
as well as the size measurement technique.
As the primary focus of this paper is not the observed size evolution,
we can only conclude that merging increases the sizes of a $10^{11}M_{\odot}$ QG by a factor of two at most from $z\sim2.5$ to 0.
While this is insufficient to explain the observed average size growth of a factor of 3-5,
it is enough to bring the average sizes of massive QGs to $1\sigma$ below the local mean stellar mass-size relation if the redshift dependence is on the milder end of the observations ($R\propto(1+z)^{-1.0}$) like in \citet{Newman2012}.
If the sizes follow a normal distribution,
the massive QGs already formed and quenched since $z\sim2.5$ evolve through merging to form the smallest 16\% (2\%) of local massive QGs since they lie at $1\sigma$ ($2\sigma$) below the mean. 
If the sizes follow a skewed distribution instead,
as shown by \citet{Newman2012},
the fraction can be even higher (e.g. up to the smallest 12.5\% for $2\sigma$ below mean following Chebyshev's inequality).
This may be a more relevant representation if these compact QGs end up to lie below the local mass-size relation, 
while the majority of later quenched QGs occupy the upper part of the relation.
Recent measurements of compact massive QGs reveal that their number densities peak at $z\sim1.8$, and decrease at lower redshifts \citep{vanderWel2014, vanDokkum2014},
therefore they must undergo structural changes.
Incidentally this is the same redshift range in which our merger rate peaks (major: $z\sim0.7-1.7$, minor: $z\sim0.7-1.5$, see Fig.~\ref{fig:merger_rates}).
We will further the discussion on the observed size evolution in Section~\ref{sec:ETG_evolution}.

Even though there may be a significant number of minor mergers rejected by the stellar mass ratio criterion (flux ratio between 1:1 and 10:1, but stellar mass ratio more extreme than 10:1),
these mergers are more likely to have non-negligible gas mass and more dissipation so it does not help to solve the problem of the observed size evolution.
The gas content of merging galaxies may explain the scatter of the redshift-size evolution \citep{Khochfar2006}.
However without gas measurements we are not able to test this hypothesis at this point.

The virial theorem predicts that equal-mass mergers do not change the stellar velocity dispersion $\sigma_{\star}$,
and minor mergers reduces the $\sigma_{\star}$ by $\sigma_{\star, 1+2}^{2} / \sigma_{\star, 1}^{2} \approx M_{1} / M_{1+2}$ if the satellite has a $\sigma_{\star}$ much lower than the massive galaxy it is merging with \citep{Bezanson2009, Naab2009}.
Using the stellar mass accretion rate we estimate that 4:1-10:1 minor mergers can only reduce the $\sigma_{\star}$ of massive galaxies by 6\% from $z=2.5$ to 0.1.
If we relax the assumption and allow 1:1 - 4:1 mergers to be equally efficient in reducing $\sigma_{\star}$,
the total stellar mass accreted implies that the $\sigma_{\star}$ decreases by maximum 25\% from $z=2.5$ to 0.1.
From this we conclude that merging is insufficient to reduce the high $\sigma_{\star}$ ($\sim 300$ km s$^{-1}$) observed in $z\sim2$ QGs \citep{Toft2012} by $\sim60\%$ to match the average of the local population.
This is consistent with claims that the addition of lower $\sigma_{\star}$ galaxies to the quiescent population at later times contribute to the decreasing average $\sigma_{\star}$ of the overall massive QG population \citep{Bezanson2012, Bezanson2013}.
We note that if a significant amount of dark matter is accreted by these massive QGs,
the total mass increases and therefore the velocity dispersion and the sizes may change without any observable stellar mass growth.

\subsection{The major merger contribution to the formation of ``new'' massive galaxies} \label{sec:numdens_evolution}

\begin{figure}[!htb]
	\centering
	\includegraphics[angle=0,width=0.5\textwidth]{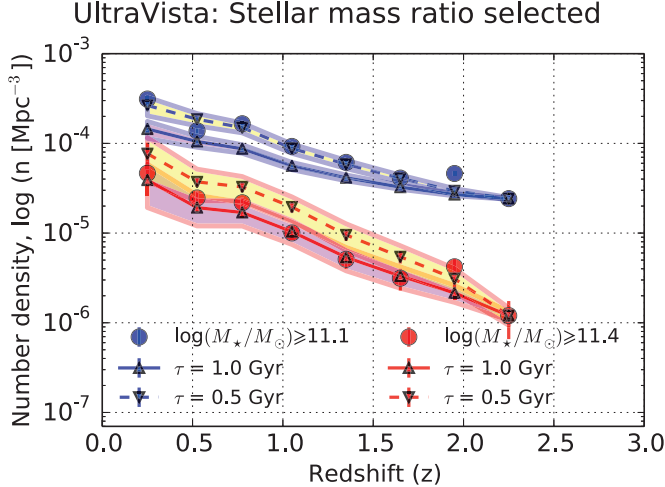}
	\caption{
	The number density evolution of the most massive galaxies ($n_{mas}$) above two stellar mass thresholds.
	The blue (red) filled circles represent the observed $n_{mas}$ of massive galaxies of log$(M_{\star}/M_{\odot}) \geqslant 11.1$ (11.4),
	and the error bars represent the Poisson error of the number counts.
	The triangles represent the major-merger driven $n_{mas}$ growth using two merger observability timescales ($\tau_{obs}=$0.5 Gyr: downward triangles, dashed lines and yellow shades; 1.0 Gyr: upward triangles, solid lines, purple shades).
	The colored shades show the uncertainty on $n_{mas}$ propagated from the Poisson errors of the number of mergers.
	The predicted major-merger driven $n_{mas}$ growth accounts for the formation of ``new'' massive galaxies above the threshold due to major merging,
	as well as the reduction in numbers of massive galaxies that merge with each other (a minor effect as observed).
	The predicted growth is normalised to the observed $n_{mas}$ of massive galaxies $z\sim2.25$ to which we are complete for major mergers.
	We only perform this exercise on the UltraVISTA catalog,
	because the 3DHST+CANDELS contain too few galaxies above these stellar mass thresholds for meaningful $n_{mas}$ constraints.
	We find that the slope of the observed $n_{mas}$ evolution of the most massive galaxies follows the predicted slope due to major merging,
	if the $\tau_{obs}\simeq 0.6-0.7$ Gyr \citep{Lotz2010a} for major merging which is roughly the average of the two timescales shown.
	To keep the slope consistent with the observed number densities,
	a maximum of $15\%$ stellar mass can be added in addition to major merging,
	implying $\leqslant6\%$ for mechanisms other than major and minor merging.
	}
	\label{fig:numdens_growth}
\end{figure}

To understand what the merger rates from Section~\ref{sec:compare_merger_rates} imply for the overall galaxy evolution,
in this section we aim to quantify the contribution of merging to the observed increase in the number density ($n_{mas}$) of massive galaxies in the redshift range $z=0.1-3$.
As shown in Section~\ref{sec:mass_growth}, 
most of the stellar mass accreted is through major merging,
so in this section we only consider major merging for which our samples are complete to higher redshifts.
Merging can affect the number counts of massive galaxies in two counteracting ways.
On one hand, 
merging among lower mass galaxies can increase the number of massive galaxies above a stellar mass threshold.
On the other hand, 
merging among massive galaxies already above the mass threshold will lead to a decreased number count.
We denote $\Delta N_{+}$ as the number of mergers with individual stellar masses lower than a given threshold,
but with the sum of their stellar masses above the threshold \citep{Robaina2009, Man2012},
and $\Delta N_{-}$ as the number of mergers with the individual stellar masses of both galaxies above the threshold.
The net change of $n_{mas}$ due to major merging is $\Delta n_{mas} (z) = (\Delta N_{+} (z) - \Delta N_{-} (z) ) / V_{comoving}(z) \times \Delta t(z) / \tau_{obs}$, 
where $V_{comoving}(z)$ and $\Delta t(z)$ are the comoving volume and the elapsed time of the redshift range,
and $\tau_{obs}$ is the merger observability timescale given the projected separation ($R_{proj}$) range.
The $\tau_{obs}$ for major mergers with $R_{proj}$ = 10-30kpc $h^{-1}$ is about 0.6-0.7 Gyr \citep{Lotz2010a} with an error of $\sim 0.4$ Gyr.
In this exercise we show the results of two values of $\tau_{obs}$ (0.5 and 1.0 Gyr).
Since we assume that no new stars are formed during mergers for the reasons discussed in Section~\ref{sec:mass_growth}, 
the presented quantities mark the minimum merger contribution to the formation of new massive galaxies.

We present the results in Figure~\ref{fig:numdens_growth}.
We find that major merging alone can explain the $n_{mas}$ evolution of galaxies more massive than $10^{11.1} M_{\odot}$ if $\tau_{obs}$ lies between 0.5 - 1 Gyr.
If $\tau_{obs}$ was systematically much longer than 1 Gyr, 
then additional mechanisms may be required to explain the $n_{mas}$ evolution of these very massive galaxies.
We note that 3DHST+CANDELS is inadequate for tracing the $n_{mas}$ growth of the most massive galaxies.
The volume probed is too small leading to large cosmic variance on the observed number density and therefore is not shown.

Taking our results further,
we use the observed $n_{mas}$ evolution of the most massive galaxies to constrain the upper limit of the stellar masses that can be added in addition to major merging.
We increase the stellar masses of all the galaxies by an arbitrary factor,
and count the number of galaxies $\Delta N_{+}^{'}$ that cross the given mass thresholds.
Its contribution to the $n_{mas}$ evolution is $\Delta N_{+}^{'} (z)/V_{comoving}(z)$.
We find that the observed $n_{mas}$ evolution is marginally consistent with a maximum 15\% of stellar mass growth of the overall massive galaxy population in addition to major merging since $z\sim2.5$.
Any non-major merging stellar mass growth beyond $15\%$ would overproduce the number of the most massive galaxies.
As shown in Section~\ref{sec:mass_growth},
minor merging accounts for $\sim9\%$ of the stellar mass accreted.
Therefore we conclude that there remains little room ($\leqslant6\%$) for the most massive galaxies to increase their stellar masses by mechanisms other than major and minor merging, 
such as star formation or very minor mergers ($\mu>$10:1).

\section{Discussions} \label{sec:discussion}

\subsection{An emerging evolutionary scenario for massive quiescent galaxies (QGs)} \label{sec:ETG_evolution}

There are comparative studies of the possible mechanisms that can explain the size evolution \citep{Hopkins2010c, Trujillo2011, Cameron2012}.
Merging, in particular dry minor merging,
appears to be a viable means to explain the observed size and velocity dispersion evolution.
However, even when we assume that all mergers were dry (dissipationless),
the size evolution inferred from our merger fraction can only account for a factor of two of size increase from $z\sim$2.5 to 0.1.
This is marginally consistent with being $1\sigma$ below the mean stellar-mass size relation of the measurement of \citet{Newman2012},
but $>2\sigma$ compared to that of \citet{vanderWel2014}.
This necessitates additional mechanisms to explain the observed size increase for the bulk of the population.

The apparent strong size evolution may be in part due to observational effects.
Our observations indicate that massive galaxies tend to merge with galaxies with lower stellar mass-to-light ratios (see Figure~\ref{fig:MtoL_z} and Section~\ref{sec:pf_ratios}).
If the younger, bluer stars of the companion are added to the outskirts of massive galaxies consisting of older stellar populations \citep{vanDokkum2010, Hilz2012, Hilz2013},
then the half-light radius ($r_{e}$) measured in rest-frame optical bands increases.
This scenario is supported by the observed negative colour gradients \citep{vanDokkum2010, Guo2011, Szomoru2011, Gargiulo2012, Szomoru2013},
and is consistent with the observation of \citet{vanderWel2014} that the $r_{e}$ of massive galaxies are smaller when measured at longer wavelengths.
\citet{Szomoru2013} show that the half-mass radii of massive QGs are on average $\sim25\%$ smaller than the half-light radii measured from the rest-frame $g$-band.
Therefore the observed size evolution is perhaps in part due to the radial dependence of the $M_{\star}/L$.
Since the number- and mass-weighted average stellar mass ratio is $\sim$ 4:1 for the mergers in this work,
the satellites may strip off their stars at the outskirts like the 5:1 intermediate mass ratio merger simulated by \citet{Hilz2013},
lending support to merging as a viable explanation for the observed size evolution and color gradients.

It is important to distinguish between the growth of individual galaxies and the evolution of the overall population.
The number density of the massive QGs evolves with redshift,
for instance massive ($10^{11}M_{\odot}$) galaxies are 30 times more abundant at $z\sim0.95$ than $z\sim2.75$ (e.g. \citealt{Marchesini2009, Ilbert2013} and references therein, also see Section~\ref{sec:numdens_evolution}).
Therefore if larger, later quenched galaxies are continuously added to the QG population,
it may be sufficient to increase the average sizes of QGs (more details about the so-called ``progenitor bias'' in \citealt{vanderWel2009a, Carollo2013, Krogager2013}).
This assumes that the sizes of QGs are correlated with their age or time since being quenched,
a trend which is observed in some works \citep{Shankar2009, vanderWel2009a, Bernardi2010, Poggianti2013} but not in others \citep{Trujillo2011, Whitaker2012}.
Another implication is that the scatter of the size evolution is expected to increase if the progenitor bias is the sole explanation for the observed size evolution,
which contradicts the constant scatter observed \citep{Trujillo2011, Krogager2013, vanderWel2014}.
Additionally, the progenitor bias alone does not explain the disappearance of compact QGs observed at $z>2$ \citep{Belli2014a, vanderWel2014, vanDokkum2014}.
The number density of compact QGs peaks at $z\sim1.6-2.2$ and decreases towards lower and higher redshifts.
Our merger fractions (stellar mass ratio selected) peak at $z\sim1-1.5$,
and one may speculate on a causal relation between the two observations.

A fixed number density selection may provide a more direct comparison between massive QGs at $z\sim2$ and their descendants at lower redshifts \citep[e.g.][]{vanDokkum2010, Behroozi2013, Leja2013}.
If the descendants of compact massive QGs at $z\gtrsim2$ are the most compact QGs in clusters in the local Universe,
the sizes of individual QGs will only need to increase by a factor of $\sim 1.6$ \citep{Poggianti2013},
which is in good agreement with the size evolution inferred from our merger rates.

Apart from the observational effects and the progenitor bias discussed above,
alternative means to increase the sizes of individual QGs have been proposed.
Some examples include AGN and/or supernova feedback \citep{Fan2008, Fan2010}, 
adiabatic cooling via the mass loss of old stars \citep{Damjanov2009, vanDokkum2014},
and halo size evolution \citep{Posti2014}. 
It is beyond the scope of this work to draw conclusions on the relative contributions of the possible options in explaining the size evolution.
We emphasise that our results provide a strong constraint:
whichever mechanisms are responsible for the observed size evolution,
there is little room for further stellar mass to be created or added ($6\%$ at most for $z=0-2.5$) for the most massive galaxies ($M_{\star} \geqslant 10^{11.1} M_{\odot}$) in order not to over-produce the observed numbers at different redshifts.

\subsection{Merger contribution to cosmic star formation} \label{sec:cosmic_SF}

The open question of whether merging is a major contributor to the cosmic star formation history (SFH) has been tackled in different ways:
Do merging galaxies have higher star formation rates compared to isolated ones \citep{Ellison2008, Patton2011, Scudder2012, Xu2012b, Yuan2012, Patton2013, Lackner2014}?
At each epoch, are star-forming galaxies primarily mergers or isolated disks \citep{Genzel2008, Shapiro2008, FSchreiber2009, Law2009, Kaviraj2013a, Kaviraj2013b, Kaviraj2014b, Kaviraj2014a}?
These different perspectives can lead to seemingly contradictory conclusions.

Despite the apparent offset of visually identified mergers from the SFR-$M_{\star}$ relation (dubbed ``main-sequence', \citealt{Hung2013}),
merging galaxies only show disturbed morphologies for a limited time ($\sim0.3$ Gyr, e.g. \citealt{Lotz2010a}).
If the duty cycle of mergers is interpreted as the cause for the scatter of the SFR-$M_{\star}$ relation,
major mergers account for a majority of the total SF at $z\sim0.6$ \citep{Puech2014}.
\citet{Patton2013} have shown that mergers can enhance SFR to $R_{proj} \sim 150$ kpc,
and such widely separated merging galaxies are likely not identified in morphological selected samples which probe later-stage mergers.
On the other hand, 
the existence of isolated star-forming disks has been used as evidence against mergers being a contributor of cosmic SF budget based on the assumption that mergers destroy disks \cite[e.g.][]{Toomre1972}.
While mergers \textit{can} destroy disks and remain a popular explanation for bulge formation \citep{Hopkins2010a},
various works have shown that disks can reform after gas-rich mergers \citep{Hopkins2009a, Stewart2009a, Puech2012}.

Even though galaxy merging may not increase the total amount of stars formed from the available cold gas reservoir,
it can trigger starburst episodes by temporarily enhancing the star formation efficiency,
leading to faster cold gas depletion \citep{Cox2008, Torrey2012}.
Detailed studies of the SFH of individual galaxies can provide an answer to whether most stars in galaxies are formed during merging or isolated phases (continuous vs bursty SFH).
In Section~\ref{sec:pf_ratios} we have shown that using the $H$-band flux ratio to select mergers leads to an increasing merger fraction evolution,
as opposed to the flat or diminishing trend seen using stellar mass ratio selected pairs.
The former merger fraction share a similar redshift evolution as the cosmic star formation rate density (e.g. \citealt{Madau2014} and references therein) albeit with considerable uncertainties:
both rise from $z\sim0$ to $z\sim1$ and reach a plateau or increase mildly from $z\sim1$ to $z\sim2.5$.
This may be a hint that at $z\gtrsim1.5$, 
massive galaxies are primarily merging with low stellar mass ($M_{1}/M_{2} >$ 10:1) but gas-rich satellites.
These mergers are classified as major or minor depending on whether the baryon mass or stellar mass ratio is used.
When inferring the merger contribution to the cosmic star formation budget,
we need to account for these ``missing'' mergers \citep{Stewart2009b} that did not enter the stellar mass ratio selection.
Future surveys of the molecular gas mass of high-$z$ galaxies are needed to make progress on this issue.

\subsection{Future prospects} \label{sec:future}

The merger fraction of massive galaxies is $<~30\%$,
resulting in low number densities of mergers ($\sim10^{-4.5} - 10^{-6}$ Mpc$^{-3}$) at $z>2$.
As we show in Section~\ref{sec:cv},
cosmic variance is the dominant source of uncertainty in merger fraction measurement with CANDELS-sized surveys,
due to the small survey area and low source number density.
We note that the merger fractions measured from UltraVISTA and 3DHST+CANDELS yield very consistent results (see Figure~\ref{fig:pf_3dhst}),
even at the redshifts where UltraVISTA is expected to be incomplete for low surface brightness satellites.
This is due to the fact that most satellites have lower $M_{\star}/L$ ratios (see Section~\ref{sec:pf_ratios} and Figure~\ref{fig:MtoL_z}).
As long as the lower limit of $R_{proj}$ is set so that no close pairs are missed due to blending,
and the relevant observability timescales are applied for the $R_{proj}$ range \citep{Lotz2010a},
deep ground-based NIR surveys like UltraVISTA and UDS provide as accurate results as \textit{HST} surveys.
Ground-based surveys have the additional advantage of larger sample sizes,
so that the evolution can be probed in finer redshift bins with small Poisson uncertainties.
Put another way,
large area surveys are crucial to mitigate cosmic variance and Poisson uncertainties in galaxy merger fraction measurements.
A limitation of the pair selection is that a minimum $R_{proj}$ must be imposed to match the resolution of the imaging data,
for example 10 kpc $h^{-1}$ in this work.
If the scientific interest is on the incidence of late stage mergers of $R_{proj}\leqslant 10$kpc $h^{-1}$ among AGNs or ULIRGS \citep[e.g.][]{Kartaltepe2010, Treister2010, Silverman2011, Kartaltepe2012, Treister2012, Ellison2013},
alternative merger identifications may be a more appropriate choice \citep[e.g.][]{LFevre2000, Conselice2003, Lotz2008a, Bluck2012, Lackner2014}.

Photometric redshifts (photo-$z$'s) are essential in removing line-of-sight projected pairs from merger samples.
The projected pair fraction is redshift dependent and can reach $N_\mathrm{projected}/N_\mathrm{mergers} \simeq 400\%$ at $z\geqslant 2$ (see Table~\ref{table:pf_massratio}).
Statistical simulations can provide an estimate for the number of projected pairs,
however photo-$z$'s are crucial for selecting real mergers for spectroscopic follow-up.
One may expect photometric samples of mergers to include more mergers due to the larger uncertainties of photo-$z$'s than spec-$z$'s,
however the merger fractions presented in this work using photometrically selected mergers are in agreement or even lower than those using spectroscopic selected mergers \citep{dRavel2009, dRavel2011, LSanjuan2011, LSanjuan2012, Tasca2014}.
Aside from the variations of the parent sample as discussed in \citet{Lotz2011},
this may be an indication that the selection effects associated with the spectroscopic merger samples outweigh the uncertainties of photo-$z$'s in photometric merger samples,
e.g. mass-incompleteness (due to flux-limited selection), 
slit/fiber placement incompleteness, 
limited sample sizes and so on.
Therefore we argue that large-area ($\gtrsim 1$ deg$^{2}$) surveys with accurate photo-$z$'s currently provide the most time-efficient datasets for measuring galaxy merger fractions.

On the theoretical front, the merging probability of galaxy pairs in close physical separations need to be quantified as a function of redshift and environment,
as discussed in Section~\ref{sec:method}.
It is also important to understand how galaxy fly-bys can impact the structure and dynamics of massive galaxies.
These are subtle yet crucial quantities that fold into the interpretation of the inferred galaxy merger rates,
which are paramount in determining whether galaxy merging is a significant driver of its evolution.


\section{Conclusions} \label{sec:conclusions}

We present the largest sample of photometrically selected mergers at $z=0.1-3$ from mass-complete catalogs,
using complementary datasets of a large area ground-based survey (UltraVISTA) and a deep spaced-based survey (3DHST+CANDELS).
We measure the galaxy major and minor merger fractions ($f_{major}$ and $f_{minor}$).
Applying the merging observability timescale ($\tau_{obs}$) from \citet{Lotz2010a},
we infer the merger rates, 
as well as the evolution in stellar mass, size and number density for massive galaxies.
We summarise our findings as follows:
\begin{enumerate}

\item 
The merger fraction shows a steep increase from $z\sim0$ to 1, 
with $f_{major}$ showing a stronger evolution than $f_{minor}$.
Using the stellar mass ratio selection (Figure~\ref{fig:pf_massratio}, left),
$f_{major}$ and $f_{minor}$ show a plateau at $z\sim1-1.8$ and diminishes beyond $z\sim1.8$.
If the observed $H$-band flux ratio selection is used instead (Figure~\ref{fig:pf_fluxratio}, right),
$f_{major}$ and $f_{minor}$ increase monotonically with redshift.
The UltraVISTA and 3DHST+CANDELS show discrepant results at $z>1.5$ due to the magnitude limit of the UltraVISTA DR1 survey.

\item Selecting mergers by the observed $H$-band flux ratio leads to an increasing merger fraction with redshift, 
while selecting mergers by stellar mass ratio shows a diminishing redshift dependence.
This variation in merger selection technique is the cause of the discrepant merger fraction measurements at $z>1.5$ in the literature \citep{Bluck2009, Williams2011, Man2012, Newman2012}.
The discrepancy is a consequence of the $M_{\star}/L$ evolution of galaxies with redshift: 
at high redshifts and lower $M_{\star}$, 
galaxies have higher star formation rates and lower $M_{\star}/L$ ratios.
The two selections produce consistent merger fractions at $z<1.5$,
but the fractions diverge at $z>1.5.$
The $H$-band flux ratio selection is biased towards bright, star-forming low-mass satellites at $z\gtrsim1.5$,
and the stellar mass ratio selection is biased against low-mass satellites which have significant cold gas mass.
Cold gas measurements for massive galaxies and their satellites are required to refine the merger definition using the baryon mass ratio.

\item Our inferred merger rates using the stellar mass ratio selection is consistent with the gas-poor ($f_{gas}<20\%$) merger rates of the simulations of \citet{Hopkins2010a}.
On the other hand, our inferred merger rates using the $H$-band flux ratio selection is consistent with their predicted gas-rich ($f_{gas} \geqslant20\%$) ones.

\item We get consistent merger rates when mergers are selected from different $R_{proj}$ bins (5-20, 10-30, 10-50, 10-100 kpc $h^{-1}$) when the relevant $\tau_{obs}$ from \citet{Lotz2010a} are applied.
However, we note that the widest $R_{proj}$ are systematically higher than the other bins,
with a more noticeable discrepancy at lower redshift.
This is consistent with the pairs at 50-100 kpc $h^{-1}$ probing large-scale structure formation.

\item The results imply that an average massive ($M_{\star} \geqslant 10^{10.8} M_{\odot}$) galaxy experiences $\sim1.0\pm0.2$ major and $\sim0.7\pm0.1$ minor mergers over the redshift range of $z=0.1-2.5$, if mergers are selected by stellar mass ratio. 
There may be an additional $\sim0.5$ major merger and $\sim0.3$ minor merger if mergers are selected by the $H$-band flux ratio.

\item The mass-weighted average stellar mass ratio is $\sim$ 3:1-4:1,
implying that the inferred stellar mass accretion rate is primarily driven by intermediate mass ratio mergers up to $z\sim2.5$.
This work extend the expectations from $z\lesssim1$ to $z\sim2.5$ that major merging is the dominant process for stellar mass accretion for massive galaxies.

\item Major and minor merging combined can at most increase the sizes by a factor of two from $z=2.5$ to 0.1 for an average $M_{\star} \simeq 10^{11} M_{\odot}$ quiescent galaxy,
if we assume that all mergers are dry. 
Additional mechanisms are thus required to explain the strong observed size evolution (factor of $\sim3-5$).

\item The observed amount of major merging is sufficient to explain the evolution of the formation of new massive ($M_{\star} \geqslant 10^{11.1} M_{\odot}$) galaxies by number density arguments.
These very massive galaxies can only increase their stellar masses by at most $\sim6\%$ during $z=0-2.5$ by processes in addition to major and minor merging,
in order to match the observed number density evolution.
This hints that star formation and very minor merging are unlikely mechanisms responsible for the observed size evolution.

\end{enumerate}

\acknowledgments{}
We are grateful for the contributions of the UltraVISTA, COMSOS, CANDELS and 3DHST collaborations for making the catalogs available for public use.
AM acknowledges Tomo Goto, Knud Jahnke, and Jennifer Lotz for helpful conversations at the early phase of this project.
AM also thanks Bo Milvang-Jensen for clarification of the UltraVISTA data, 
and Anna Gallazzi for clarifying the stellar population models.
The Dark Cosmology Centre is funded by the Danish National Research Foundation.
We acknowledge the HPC facility at the University of Copenhagen for providing the computing resources used in this work.
ST and AZ gratefully acknowledge support from the Lundbeck Foundation.

This work has made use of the UltraVISTA catalog,
which is based on data products from observations made with ESO Telescopes at the La Silla Paranal Observatory under ESO programme ID 179.A-2005 and on data products produced by TERAPIX and the Cambridge Astronomy Survey Unit on behalf of the UltraVISTA consortium.

This work is in part based on observations taken by the 3D-HST Treasury Program (GO 12177 and 12328) with the NASA/ESA HST, which is operated by the Association of Universities for Research in Astronomy, Inc., under NASA contract NAS5-26555.
The 3DHST+CANDELS catalog is compiled using the datasets in these papers:
\citet{Dickinson2003, Steidel2003, Capak2004, Giavalisco2004, Erben2005, Hildebrandt2006, Taniguchi2007, Barmby2008, Furusawa2008, Wuyts2008, Erben2009, Hildebrandt2009, Nonino2009, Cardamone2010, Retzlaff2010, Grogin2011, Kajisawa2011, Koekemoer2011, Whitaker2011, Bielby2012, Brammer2012, Hsieh2012, McCracken2012, Ashby2013}; Almaini/Foucaud in prep and Dunlop et al. in prep.

\appendix
\section{Could we be missing mergers?} \label{sec:missing_mergers}

In order to measure the merger fraction evolution robustly,
it is essential to ensure completeness in the identification of merging satellites especially at high redshifts.
We assess the completeness of faint satellites in two aspects:
\begin{enumerate}
  \item Stellar mass completeness: is UltraVISTA mass complete at high-$z$ for the 10:1 satellites?
  \item Surface brightness (SB): do we miss low SB faint satellites?
\end{enumerate}
We present our analysis in the following subsections.

\subsection{Stellar mass completeness} \label{sec:mass_completeness}

\begin{figure}[h!]
	\centering
	\includegraphics[angle=0,width=0.5\textwidth]{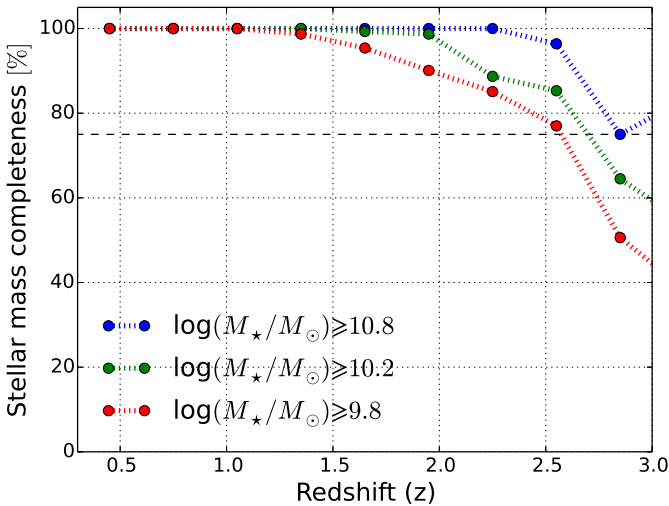}
	\caption{
	This plot shows the stellar mass completeness of the UltraVISTA catalog as a function of redshift.
	The stellar mass completeness here is computed by comparing the $K$-band magnitude distribution of UltraVISTA to the deeper FIREWORKS catalog.
	The stellar mass bins of the massive galaxies of log$(M_{\star}/M_{\odot}) \geqslant 10.8$,
	as well as their 4:1 and 10:1 satellites are shown in different colors as indicated in the legend.
	The dashed line shows the $75\%$ completeness limit.
	}
	\label{fig:mass_completeness}
\end{figure}

We estimate the stellar mass ($M_{\star}$) completeness of the UltraVISTA catalog by comparing the $K$-band magnitudes and photo-$z$'s of the detected galaxies with those of the deeper $K$-band selected FIREWORKS catalog ($K = 24.3$ at 5$\sigma$ depth, \citealt{Wuyts2008}) in the Chandra Deep Field South.
Assuming that the FIREWORKS catalog is 100\% complete, 
we take the fractions of massive galaxies in FIREWORKS above different $M_{\star}$ in different redshift bins which are fainter than the UltraVISTA survey magnitude limit as the mass completeness limits.
The results are shown in Figure~\ref{fig:mass_completeness}.
From this comparison we estimate that for the UltraVISTA sample,
massive galaxies of log$(M_{\star}/M_{\odot}) \geqslant 10.8$ are $>75\%$ complete at $z\leqslant3$.
Major ($\mu \geqslant$ 4:1) satellites of log$(M_{\star}/M_{\odot}) > 0.25 \times$ log$(10.8) = 10.2$ are above 80\% complete for $z\leqslant2.7$.
Minor satellites (4:1 $\leqslant \mu \leqslant$ 10:1) of log$(M_{\star}/M_{\odot}) > 0.1 \times$ log$(10.8) = 9.8$ are above 80\% complete for $z\leqslant2.4$.
We list the $>75\%$ limits in Table~\ref{table:completeness_all}.

The CANDELS survey is sensitive to faint objects ($H=26.9$ at 5$\sigma$ depth, \citealt{Grogin2011}).
For example quiescent galaxies with $M_{\star} = 10^{10} M_{\odot}$ are 50\% complete at $z\sim2.8$ (3.2) for wide and deep regions \citep{Guo2013},
therefore we expect the stellar mass completeness not to be an issue.
\capstartfalse
\begin{deluxetable}{ccc} 
	\tablecolumns{3}
	\tablewidth{0pc}
	\tablecaption{Completeness limits}
	\tablehead{   
	  \colhead{} &
	  \colhead{Mass completeness$>75$\%} &
	  \colhead{SB complete limit}
	}
	\startdata
	
	\cutinhead{Massive galaxies / logM=10.8}
	UltraVISTA & $z=3.3$ & $z>3.5$ \\
	CANDELS & \nodata & $z>3.5$ \\ 

	\cutinhead{1:4 satellites / logM=10.2}
	UltraVISTA & $z=2.7$ & $z=2.4$ \\ 
	CANDELS & \nodata & $z=3.0$ \\ 
	
	\cutinhead{1:10 satellites / logM=9.8}
	UltraVISTA & $z=2.4$ & $z=1.5$ \\
	CANDELS & \nodata & $z=2.5$ \\

	\enddata
	\tablecomments{
	We tabularise the redshift and resolution limits to which we are complete for even the faintest low surface brightness galaxies (maximally old stellar population) of the given stellar masses for a range of Sersic profiles.
	The second column are the redshift limits for stellar mass completeness of $>75\%$ derived by comparing UltraVISTA galaxies to the deeper FIREWORKS catalog, 
	as described in Appendix~\ref{sec:mass_completeness}.
	The third column shows the Sersic tested SB limits derived by simulating maximally old galaxies of a range of Sersic profiles, 
	as detailed in Appendix~\ref{sec:sb_limit}.
	We list the redshift to which the catalogs are complete to the detection of such galaxies with the Sersic profiles simulated.
	}
	\label{table:completeness_all}
\end{deluxetable}
\capstarttrue


\subsection{Surface brightness limits: Modelling the faintest possible satellites} \label{sec:sb_limit}
The detection of objects at faint magnitudes is sensitive to their surface brightness (SB) profiles and the source extraction thresholds.
In order to test the redshift limit up to which we are complete to detecting the faintest possible satellites,
we simulate the source detection by simulating galaxies with a range of Sersic profiles with magnitudes determined by a dust-free, maximally old stellar population at given $M_{\star}$ and redshifts.
The effective half-light radii ($r_{e}$) assumed are the extrema calculated from the observed scaling relations and/or simulations,
as described in detail below.
To emulate the actual observations of UltraVISTA and CANDELS,
the Sersic profiles are smoothed to the instrument PSF and added to images with blank patches of sky,
and SExtractor is run with the object detection settings of the respective catalogs.

Structural measurements of intermediate mass ($M_{\star} \sim 10^{9.8}M_{\odot}$) galaxies at $z\gtrsim2$ are sparse due to their faintness.
We list the possibilities here and select the extreme sizes for our simulations.
\begin{enumerate}
  \item 
Observationally,
the sizes of local elliptical or early-type galaxies scale with stellar mass as $R \propto M_{\star}^{0.5-0.56}$ for $M_{\star} > 10^{10.6} M_{\odot}$.
The observed $z\sim2$ stellar mass-size relation has a similar slope \citep{Williams2010, Newman2012}.
  \item 
If intermediate mass galaxies have the same stellar density as massive galaxies,
then the radius scales with stellar mass as $R \propto M_{\star}^{1/3}$. 
  \item 
Lastly,
numerical simulations for merger-driven size evolution have shown that a Hernquist profile in projection can be described by a Sersic index of $n\sim2.6$ \citep{Hilz2012, Hilz2013}. 
These simulations use the same scale radius for the stellar halos of the host galaxy and the satellite which has only a tenth of the host stellar mass for the ``diffuse'' case.
\end{enumerate}

\subsubsection{Sizes}
Considering the above mentioned possibilities, 
we simulate the two extreme sizes of a $M_{\star}=10^{9.8} M_{\odot}$ quiescent galaxy: 
the most compact (constant stellar density: $R \propto M_{\star}^{1/3}$) and the most extended (simulation: $R_{massive} = R_{intermediate}$).
Observations show that a $M_{\star}=10^{10.8} M_{\odot}$ quiescent galaxy has log$(r_{e}/kpc) \sim 0.2$ at $1.5 < z < 2$ and log$(r_{e}/kpc) \sim 0.04$ at $2 < z < 2.5$ \citep{Williams2010, Newman2012, Cassata2013}, with a scatter of $\sigma_{log(r_{e}}) \sim 0.25$.
We scale the sizes to one-tenth of the stellar mass with the extreme scenarios,
e.g. our simulated $M_{\star}=10^{9.8} M_{\odot}$ galaxy at $z=2.5$ has $r_{e}$ of 0.29 kpc (compact) to 1.95 kpc (extended), 
equivalent to 0.035$\arcsec$ and 0.248$\arcsec$.
\subsubsection{Magnitudes}
We assume a maximally old, dust-free stellar population with a single burst and highest metallicity ($Z$=0.03) to compute the faintest possible magnitudes for these intermediate mass galaxies using the updated version (2012) of the stellar population synthesis model library of \citet{Bruzual2003}.
This corresponds to magnitude limits of $H=26.34$ and $K=25.41$ for a $M_{\star}=10^{9.8} M_{\odot}$ maximally old galaxy at $z=2.5$.
\subsubsection{Other assumptions}
We simulate different light profiles using three Sersic indices ($n=$[0.5, 1, 4]),
in which the latter two represent the exponential disk profile and the de Vaucouleurs profile respectively.
We assume two axial ratios of $q=[0.5,1]$, 
though we note that lower axial ratios are easier to detect when the source is closer to the SB limit.

\subsubsection{Method and results}
With the assumed parameters we generate Sersic models according to the $H$ and $K$ limits.
We smooth the images with a Gaussian beam corresponding to the PSF size of the imaging surveys.
Then we add them to blank regions on the CANDELS-wide $H$-band and the UltraVISTA $K$-band images,
and we extract sources from the simulated images with the corresponding SExtractor settings of the two surveys.

We outline the results of our simulation for both catalogs.
For the UltraVISTA DR1 catalog,
as long as the source is brighter than $K$=24.2-24.3 mag arcsec$^{2}$, 
we are able to extract the sources for all the Sersic models simulated.
This corresponds to $z=2.4$ (1.5) for using UltraVISATA DR1 to detect major (minor) satellites.
The limit for the CANDELS wide catalog is $H$=26.45 mag arcsec$^{2}$, 
corresponding to $z=3$ (2.5) for major (minor) satellites.
We note that these limits are more constraining that those derived from a simple stellar mass completeness argument (Appendix~\ref{sec:mass_completeness}).

From this test we observe that the source detection for faint objects close to the SB limit depends on the following structural parameters:
(1) $r_{e}$:
for a given integrated magnitude,
the larger the $r_{e}$ the lower the SB per pixel. Sufficient pixels (10 pixels following UltraVISTA and CANDELS settings) above the SB threshold are required for a detection;
(2) $n$: for a given integrated magnitude,
the combination of a very low $n$ and very extended $r_{e}$ may lead to too low SB/pix for detection.
On the other hand, 
for a very high $n$ and very compact $r_{e}$ a non-detection may result due to the insufficient number of pixels above the detection threshold;
(3) $q$: if the axis ratio is close to 1, the flux densities are divided over more pixels than the case of a lower $q$, resulting in an insufficient number of pixels above the detection threshold.

We note that our derived limits may be subject to change,
if there are systematic uncertainties in the magnitudes and/or the stellar mass.
Namely, the magnitude limits are derived from dust-free models, 
which may be reasonable assumptions given that the faintest possible galaxies at $z=2.5$ are not actively star-forming.
On the other hand, 
there are known systematic uncertainties in stellar masses ($\sim 0.2$ dex) and ages from SED fitting due to different assumptions of IMF or stellar population synthesis model.
If the modeled magnitudes are actually fainter or if the stellar masses are underestimated,
then our SB completeness limit may be lower than the numbers quoted here.


\section{Comparison with other merger fraction studies} \label{sec:pf_lit}

We only compare our results with previous merger fraction measurements using the close pair selection but not the morphological selection \citep[e.g.][]{LFevre2000, Conselice2003, Lotz2008a, Heiderman2009, Jogee2009, Bluck2012}.
As the morphological selection is sensitive to the imaging quality,
merger fraction measurements may be subject to large systematic uncertainties beyond $z\sim1$.
We refer readers to \citet{Lotz2011} for a comprehensive review on the two methods,
and focus on comparing our results with works that use the close pair method to identify mergers.
We note that for the few studies which cover a different $R_{proj}$ range than our data points shown on Figure~\ref{fig:pf_lit},
we use the observability timescales of \citet{Lotz2010a} to correct the merger fractions for a fair comparison.

\subsection{Merger fraction at $z\geqslant1.2$}

We compare our merger fractions with $z\geqslant1.2$ studies using the close pair selection.
As the selection criteria vary slightly across studies, we re-run our selection according to the published studies for a fair comparison.

\begin{figure*}[!Htp]
	\begin{minipage}[b]{0.56\linewidth}
	\centering
	\includegraphics[angle=0,width=\textwidth]{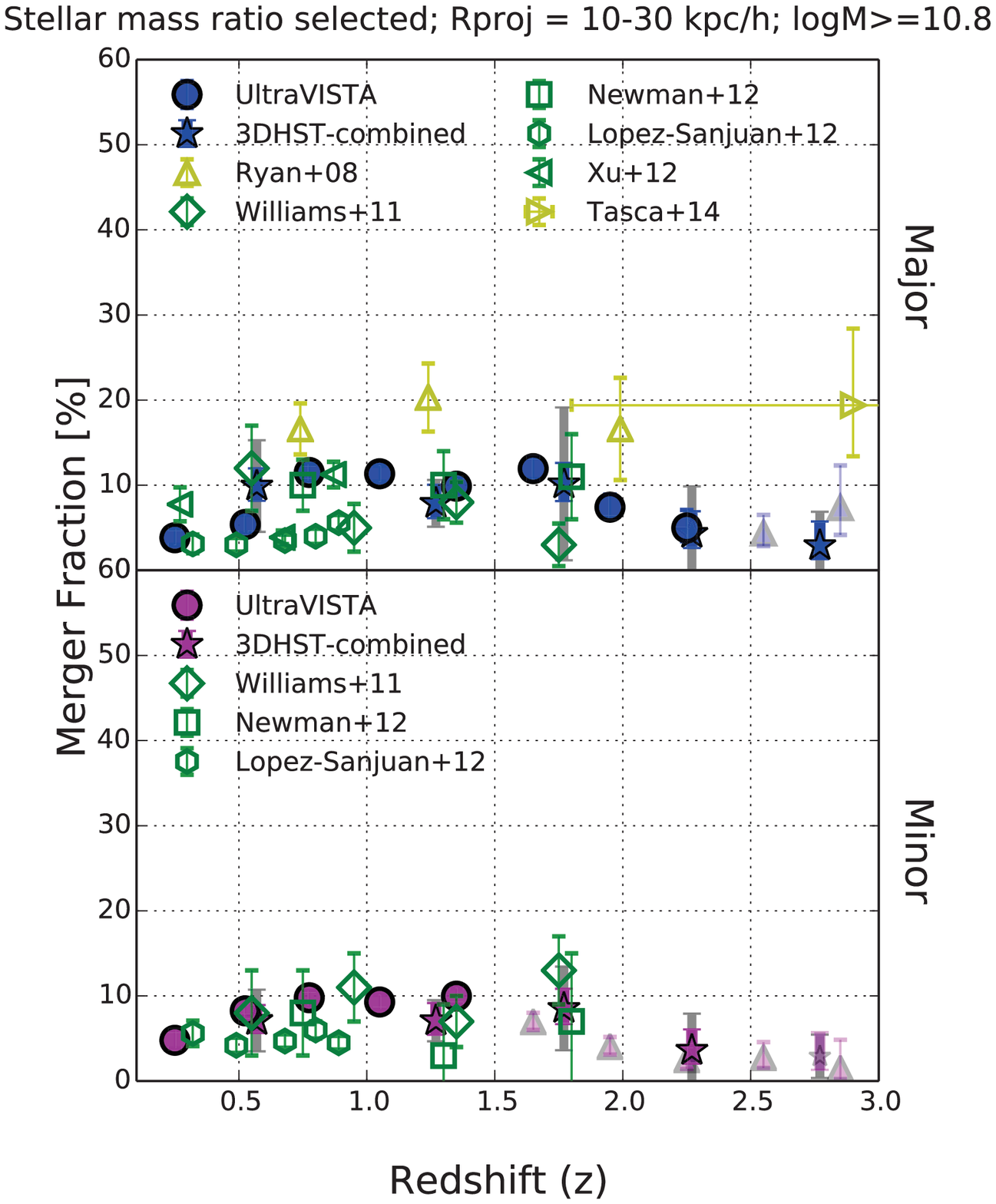}
	\label{fig:pf_lit_massratio}
	\end{minipage}
	\begin{minipage}[b]{0.5\linewidth}
	\centering
	\includegraphics[angle=0,width=\textwidth]{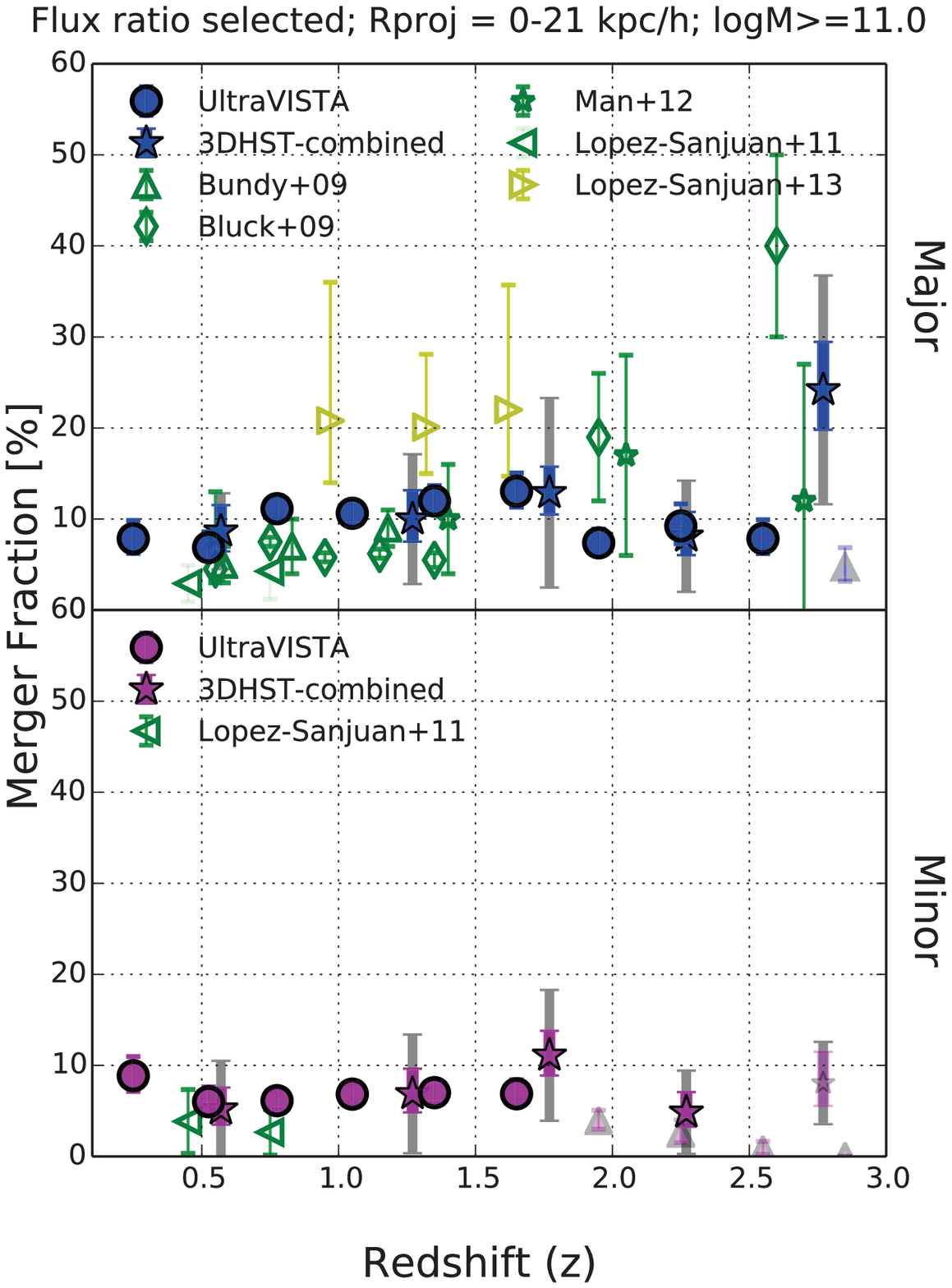}
	\label{fig:pf_lit_fluxratio}
	\end{minipage}
	\caption{
		A comparison of the major (blue filled symbols) and minor (magenta) merger fractions presented in this work with other close pair studies using similar selection criteria.
		The literature points are plotted in green / yellow open symbols.
		The green points represent the works directly comparable to our stellar mass range and the yellow points show the works concerning mergers around less massive galaxies (log$(M_{\star}/M_{\odot})<10.5$).
		We use the stellar mass ratio selection (left) and the $H$-band flux ratio selection (right) following the two methods for the respective works, as described in Section~\ref{sec:pf_ratios}.
		\textbf{\textit{Left:}} Stellar mass ratio selected mergers \citep{Ryan2008, Williams2011, LSanjuan2012, Newman2012, Xu2012a, Tasca2014}. 
		We search for satellites within $10 \leqslant R_{proj} \leqslant 30$ kpc $h^{-1}$ around massive quiescent (log$(M_{\star}/M_{\odot}) \geqslant 10.8$ and log(sSFR)$<$-10.7) galaxies.
		\textbf{\textit{Right:}} $H$-band flux ratio selected mergers \citep{Bundy2009, Bluck2009, LSanjuan2011, Man2012, LSanjuan2013}:
		We search for satellites with $R_{proj} \leqslant 30$ kpc ($R_{proj} \leqslant 21$ kpc $h^{-1}$) around massive (log$(M_{\star}/M_{\odot}) \geqslant 11$) galaxies using the $H$-band flux ratios.
		As we show in Section~\ref{sec:pf_ratios},
		the ``major'' flux ratio selection includes ``minor'' stellar mass ratio mergers at $z>1.5$.
		This leads to an increasing merger fraction at $z>1.5$.
		There are slight differences among the merger selections as described in Appendix~\ref{sec:pf_lit}.
		We find consistent conclusions with these studies that the major and minor merger fractions are flat or even diminishing when the stellar mass ratio selection is used,
		but an increasing trend is observed at $z>1.5$ when the flux ratio is used.
		}
	\label{fig:pf_lit}
	
\end{figure*}

We compare our merger fractions with similar studies that select mergers using the stellar mass ratio \citep{Williams2011, Newman2012}
\footnote{In the case of \citet{Newman2012}, 
we convert their mass limit from a Salpeter IMF to a Chabrier IMF to match this study.}.
In these studies,
the projected separation limits are $R_{proj}$ = 13-30 kpc $h^{-1}$ and 10-30 kpc $h^{-1}$ respectively.
We replicate the selections by slightly modifying our criteria:
we search for satellites around massive quiescent galaxies ($M_{\star} \geqslant 10^{10.8}M_{\odot}$ and sSFR $< 10^{-10.7}$),
using a limit of $R_{proj}$=10-30 kpc $h^{-1}$.
We note that the results of \citet{Newman2012} are based on satellites around quiescent galaxies at lower stellar masses ($M_{\star} \geqslant 10^{10.5}M_{\odot}$).
We check that lowering the stellar mass cut by 0.3 dex gives consistent merger fractions within the large Poisson uncertainties,
as is also shown in \citet[Table~3]{Newman2012}.
The comparison is shown in Figure~\ref{fig:pf_lit_massratio} (left).
We find our $f_{major}$ to be consistent with that of \citet{Newman2012},
and the one measured from UltraVISTA is $\sim 1-2\sigma$ higher than that from \citet{Williams2011} at $z\sim1$ and 1.8.
We note that in these redshift bins,
\citet{Williams2011} show slightly higher $f_{minor}$ than in other fields.
Therefore we conclude that the combined $f_{major}$ and $f_{minor}$ measured in our data and in \citet{Williams2011} are in good agreement.
The discrepancy of $\sim3\%$ in the $f_{major}$ can be explained by the separation of major and minor mergers,
as well as cosmic variance and photo-$z$ criterion variation.
This discrepancy does not affect the conclusions made in this work.

\citet{Bluck2009} and \citet{Man2012} search for satellites of $H$-band flux ratios down to 4:1 around galaxies more massive than $10^{11} M_{\odot}$,
within projected separations of $R_{proj} \leqslant 30$ kpc, i.e. 21 kpc $h^{-1}$.
In particular, \citet{Bluck2009} impose a lower limit of $R_{proj} > 5$ kpc to screen out confused pairs which are likely unresolved with NICMOS.
This comparison is illustrated in Figure~\ref{fig:pf_lit_fluxratio} (right).
Our $f_{major}$ is consistent with these studies.

\citet{Ryan2008} present the first measurement of the $f_{major}$ at $z>1$ in the HUDF 
using the stellar mass ratio selection.
They use a smaller $R_{proj} \leqslant$ 20 kpc $h^{-1}$ and search for satellites around galaxies of $10^{10} M_{\sun}$, which is six times lower than our mass criteria.
This may explain why their $f_{major}$ to be $50\%$ higher than ours.

As discussed in Section~\ref{sec:future},
flux-limited spectroscopic surveys may lead to biased merger fractions due to mass incompleteness, slit/fiber collision, etc.
Bearing in mind the difference in the merger selection,
we compare our results using photometric mergers with those using spectroscopic mergers.
Our results are consistent with \citet{LSanjuan2011} who measure the $f_{major}$ and $f_{minor}$ of $\gtrsim L_{B}^{\star}$ galaxies from the spectroscopic survey of VVDS up to $z\sim1$ using the $B$-band flux ratio selection.
The observed $H$-band corresponds approximately to the rest-frame $B$-band at $z\sim2.5$ and therefore our results using the flux ratio selection are directly comparable to their work.
\citet{LSanjuan2013} and \citet{Tasca2014} extend measurements of spectroscopic merger fractions to $z>1.2$,
in which the former use a flux ratio selection for star-forming galaxies and the latter a stellar mass ratio selection.
Both works report a $f_{major}$ of $15-20\%$.
Our major merger fraction are marginally consistent with that of \citet{LSanjuan2013} although we note that their primary sample consists of star-forming galaxies only,
and may include more mergers if merging does trigger star formation activity.
Our merger fractions are $\gtrsim10\%$ lower than that of \citet{Tasca2014}.
Both of these studies sample the mergers around less massive galaxies (0.8-1.7 dex lower than our mass limit),
and we speculate that it may account for the higher fractions.

\subsection{Merger fraction at $z\leqslant1.2$}

Our merger selection criteria are very similar to those of \citet{Bundy2009}, \citet{LSanjuan2012} and \citet{Xu2012a} so we detail our comparison here.

\citet{Bundy2009} select mergers photometrically with the $K$-band flux ratio,
and report a mildly increasing $f_{major}$ for massive ($>10^{11} M_{\odot}$) galaxies from $z=0$ to 1.2.
When compared to our $f_{major}$ using the $H$-band flux ratio for the similar $M_{\star}$ and $R_{proj}$ range (Figure~\ref{fig:pf_lit_fluxratio}, right), 
our results are in good agreement with theirs.

\citet{LSanjuan2012} measure the $f_{major}$ and $f_{minor}$ of massive ($>10^{11} M_{\odot}$) galaxies in zCOSMOS at $z = 0-1$,
selecting mergers by stellar mass ratio and relative velocity $\delta \nu \le 500$ km s${-1}$.
They find a redshift dependence of the $f_{major}$ as $(1+z)^{1.4}$,
and a redshift-constant $f_{minor}$ in this redshift range.
\citet{Xu2012a} present results for $f_{major}$ at $z=0-1$ for COSMOS with similar selection criteria.
We compare to their $f_{major}$ for galaxies with log($M_{\star}/M_{\odot}$)=11-11.4.
Our results are consistent to these two works, 
as shown in Figure~\ref{fig:pf_lit_massratio} (left).

\citet{Lotz2011} demonstrate that the variation in selecting the parent galaxy sample and the mass ratio probe leads to different redshift trends in the merger fraction.
Therefore we do not compare our results directly with the pair fraction measurements at $z\leqslant1.2$ with different selection criteria \citep[e.g.][]{Bundy2004, Xu2004, Bell2006, DePropris2007, Kartaltepe2007, Lin2008, McIntosh2008, Patton2008, Rawat2008, dRavel2009, Robaina2010, dRavel2011}.
We note that once the selection differences are accounted for,
the merger rate per galaxy presented in Section~\ref{sec:compare_merger_rates} of this work is consistent with those inferred from these works as presented in \citet{Lotz2011}:
both follow a monotonically increasing trend from $z\sim0$ to $z\sim1.2$.

\bibliographystyle{apj} 

\end{document}